\documentclass[aps,pra,twocolumn,superscriptaddress,floatfix,nofootinbib,showpacs,longbibliography]{revtex4-1}
\usepackage[utf8]{inputenc}  
\usepackage[T1]{fontenc}     
\usepackage[british]{babel}  
\usepackage[sc,osf]{mathpazo}
\usepackage[scaled=0.86]{berasans}  
\usepackage[colorlinks=true, citecolor=blue, urlcolor=blue]{hyperref}  
\usepackage{graphicx}
\usepackage{tabularray}
\UseTblrLibrary{amsmath}
\usepackage[babel]{microtype}  
\usepackage{amsmath,amssymb,amsthm,bm,amsfonts,mathrsfs,bbm} 

\usepackage{xspace}  
\usepackage{pgf,tikz}
\usepackage{xcolor}
\usepackage{multirow}
\usepackage{array}
\usepackage{bigstrut}
\DeclareUnicodeCharacter{020B}{\^{\i}}
\usepackage{braket}
\usepackage{color}
\usepackage{natbib}
\usepackage{multirow}
\usepackage{mathtools}
\usepackage{float}
\usepackage{xcolor,colortbl}
\usepackage{physics}
\usepackage{amsmath}
\usepackage{color}
\usepackage{bbold}
\usepackage[justification=justified, format=plain]{subcaption}
\usepackage[justification=raggedright]{caption}

\newcommand{\be}{\begin{equation}}
\newcommand{\ee}{\end{equation}}
\newcommand{\ba}{\begin{eqnarray}}
\newcommand{\ea}{\end{eqnarray}}

\newtheorem{theorem}{Theorem}

\newtheorem{example}{Example}

\def\>{\rangle}
\def\<{\langle}

\begin{document}
	
\title{Detecting genuine multipartite entanglement using moments of positive maps 
 }

\author{Saheli Mukherjee}
\email{mukherjeesaheli95@gmail.com}
\affiliation{S. N. Bose National Centre for Basic Sciences, Block JD, Sector III, Salt Lake, Kolkata 700 106, India}

\author{Bivas Mallick}
\email{bivasqic@gmail.com}
\affiliation{S. N. Bose National Centre for Basic Sciences, Block JD, Sector III, Salt Lake, Kolkata 700 106, India}

\author{Sahil Gopalkrishna Naik}
\email{sahiln1112@gmail.com}
\affiliation{S. N. Bose National Centre for Basic Sciences, Block JD, Sector III, Salt Lake, Kolkata 700 106, India}

\author{Ananda G. Maity}
\email{anandamaity289@gmail.com}
\affiliation{S. N. Bose National Centre for Basic Sciences, Block JD, Sector III, Salt Lake, Kolkata 700 106, India}
\affiliation{Networked Quantum Devices Unit, Okinawa Institute of Science and Technology Graduate University, Onna-son, Okinawa 904-0495, Japan}
\affiliation{School of Physical Sciences, Indian Institute of Technology Goa, Ponda 403401, Goa, India}

\author{A. S. Majumdar}
\email{archan@bose.res.in}
\affiliation{S. N. Bose National Centre for Basic Sciences, Block JD, Sector III, Salt Lake, Kolkata 700 106, India}

\begin{abstract}
Genuine multipartite entanglement (GME) represents the strongest form of entanglement in multipartite systems, providing significant advantages in various quantum information processing tasks. In this work, we propose an experimentally feasible scheme for detecting GME, based on the truncated moments of positive maps. Our method avoids the need for full state tomography, making it scalable for larger systems. We provide illustrative examples of both pure and mixed states to demonstrate the efficacy of our formalism in detecting inequivalent classes of tripartite genuine entanglement. We further demonstrate the detection of quadripartite genuine entanglement, underscoring the effectiveness of our method in identifying entanglement beyond the tripartite case. Finally, we present a proposal for realising these moments in real experiments.
\end{abstract}

\maketitle

\section{Introduction}\label{s1}
Quantum entanglement \cite{horodecki2009quantum}, a hallmark of non-classical correlations, plays a central role in enabling fundamental quantum information processing tasks. It serves as a key resource in diverse applications such as 
quantum communication \cite{bennett1992communication,bennett1993teleporting,bennett2002entanglement,
wolff2025fundamental}, quantum cryptography \cite{ekert1991quantum,lo1999unconditional,yin2020entanglement}, secret sharing \cite{hillery1999quantum,cleve1999share,karlsson1999quantum}, and quantum computation \cite{jozsa2003role,ding2007review}. The correlations exhibited by entangled systems are inherently quantum, with no classical analogue \cite{vogel2014unified,killoran2016converting,d2020classical}, motivating extensive efforts to characterize and harness entanglement for both foundational insights and operational advantages \cite{plenio2014introduction,bengtsson2017geometry}. At the heart of this endeavour lies the challenge of entanglement detection: the ability to determine whether a given quantum state is entangled or not is not only of deep theoretical importance but also a necessary prerequisite for realizing practical quantum advantage.

In the bipartite setting, a wide range of techniques have been developed to address this challenge \cite{guhne2009entanglement}. One of the most popular methods for detecting entangled states 
involves the usage of positive, but not completely positive maps. Among these, the transposition map plays a crucial role, leading to the development of the positive under partial transposition (PPT) criterion \cite{peres1996separability, horodecki2001separability}, which serves as a necessary and sufficient condition for entanglement detection in bipartite systems having dimension $\le 6$. 
Other well-established techniques include the reduction criterion \cite{horodecki1999reduction}, the realignment method \cite{chen2003matrix,rudolph2005further}, the range criterion \cite{horodecki1997separability,bennett1999unextendible,bruss2000construction}, and the majorization criterion \cite{nielsen2001separable}. All these methods collectively contribute to the broader framework of entanglement detection, enabling a deeper understanding of quantum correlations.

Moving beyond the bipartite case, real-world scenarios often involve networks consisting of multiple parties \cite{perseguers2008entanglement,perseguers2013distribution,navascues2020genuine,bugalho2023distributing} where entanglement distribution can be intricate. The strongest form of entanglement in a multipartite system is the genuine multipartite entanglement (GME) \cite{walter2016multipartite,horodecki2024multipartite}. GME is a crucial resource in various quantum information tasks, including communication complexity \cite{buhrman1999multiparty,chakraborty2025scalable}, quantum thermodynamics \cite{puliyil2022thermodynamic,sun2024genuine,joshi2024experimental}, and quantum key distribution \cite{epping2017multi,holz2020genuine}. Moreover, in many instances, GME gives a significant advantage over bipartite entanglement \cite{Ac_n_2007,perseguers2013distribution,yamasaki2018multipartite,
navascues2020genuine}, demonstrating its superiority in several information processing 
applications. GME has also been studied for sequential measurements, leading to interesting consequences \cite{maity2020detection,Gupta21}. The detection of GME is therefore, a crucial aspect of research in multipartite entanglement theory. 

However, the detection of GME suffers from many complications due to the complex structure of entanglement in multipartite systems. For instance, some tripartite states are separable across all bipartitions yet remain not fully separable \cite{bennett1999unextendible}, while others are entangled across all bipartitions but do not exhibit genuine multipartite entanglement \cite{horodecki1995violating, liang2014anonymous}. Despite these difficulties, several attempts have been made to detect GME \cite{guhne2009entanglement,PhysRevLett.104.210501,PhysRevA.83.062325,PhysRevA.96.052314, PhysRevX.7.021042, PhysRevLett.122.060502,Zhou_2019, PhysRevA.107.052405,PhysRevLett.133.150201}. Similar to the bipartite case, positive maps (transposition map, reduction map, etc.) can also detect GME \cite{horodecki2001separability,clivaz2017genuine, Vaishy_2022, Mallick:2024tit}. Entanglement detection using positive maps suffers from the major drawback that these maps being unphysical, cannot be directly implemented in an experimental setup. Other methods include witness-based detection schemes where the expectation value of the witness operator is positive on all biseparable states and a negative expectation value indicates the signature of GME \cite{PhysRevLett.92.087902,Chru_ci_ski_2014,PhysRevLett.113.100501}. Nevertheless, such witness operator-based entanglement detection schemes require prior information about the state.

In this work, we propose a scheme for detecting GME using truncated (a finite number of) moments of positive maps (namely, the transposition and the reduction map), as illustrated through examples of both
pure and mixed states.  Notably, the transposition and reduction maps arise from the Lindblad structure \cite{lindblad1976generators}, a well-established and physically motivated framework in open quantum system dynamics. The problem of bipartite entanglement detection via truncated moments  is well studied in the literature \cite{cieslinski2024analysing}. Protocols for experimental realization of moments are discussed in \cite{PhysRevLett.121.150503,elben2020mixed, neven2021symmetry}. Moreover, the idea of truncated moments has also been used to detect several other quantum features such as non-Markovian dynamics \cite{PhysRevA.109.022247},
non-classicality in quantum optics \cite{PhysRevA.111.032406}, and Kirkwood-Dirac nonpositivity \cite{chakrabarty2025probing}.  

The concept of using moments for NPT (nonpositive under partial transposition) entanglement detection was introduced in \cite{elben2020mixed}. This was followed by an optimal entanglement detection criterion based on the partial transpose map using Hankel matrices \cite{yu2021optimal}. But in higher dimensions, there exists undistillable entangled states that cannot be detected using partial transpose moments. In order to circumvent this, the authors in \cite{aggarwal2024entanglement} proposed a higher-dimensional entanglement detection criterion using partial realigned moments. Refs. \cite{wang2022operational,ali2025detection,nzrc-8yrt} generalized the concept of defining moments in higher-dimensional systems for different positive maps. Furthermore, the detection of $GHZ$ and $W$ state is discussed in \cite{ali2025detection}, but the proposed criterion detects the NPT-ness of the above states, and is incapable of detecting the genuine entanglement in them.

Unlike standard positive map techniques, our moment-based criterion for the identification of GME allows these moments to be experimentally measurable by realizing the expectation values of certain operators.
Our approach offers a significant advantage in terms of resource efficiency. The estimation of moments involves evaluating linear functionals, which can be experimentally implemented using shadow tomography 
\cite{aaronson2018shadow,huang2020predicting}. This is highly advantageous compared to full state tomography, which requires an exponential number of state copies, whereas a polynomial number of state copies suffices for the moment-based approaches. Moreover, our approach does not require any prior information about the state, unlike the witness operator-based detection schemes.

The paper is organized as follows. In Sec. \ref{s2}, we provide a brief overview of positive maps and bipartite entanglement detection. We discuss the formalism of generating these maps from the Lindblad structure by suitable parametrization. An introduction to GME detection and the partial transpose moment criterion for bipartite systems is also discussed here. In Sec. \ref{s3}, we provide our truncated moment-based criteria to detect GME.  We also provide explicit examples in the tripartite and quadripartite scenarios to support our detection scheme. We then provide an experimental proposal to implement moments of the transposition map in Sec. \ref{s4}. Finally. in Sec. \ref{s5}, we summarize our main findings along with some future perspectives.

\section{Preliminaries}\label{s2}
We first introduce the terminologies and notations used throughout the paper. Let $\mathbb{C}^d$ represent a finite, $d$-dimensional complex Hilbert space and $\mathcal{B}(\mathbb{C}^d)$ denote the set of bounded operators acting on it. Quantum states are positive, trace one operators known as density operators that belong to the set $\mathcal{D}(\mathbb{C}^d)$, which is a strict subset of the set of bounded operators, i.e., $\mathcal{D}(\mathbb{C}^d) \subset \mathcal{B}(\mathbb{C}^d)$. Linear maps that transform operators between Hilbert spaces are denoted as $\Lambda : \mathcal{B}(\mathbb{C}^{d}) \rightarrow \mathcal{B}(\mathbb{C}^{d})$. For composite systems, the entire state space can be classified into two categories- separable and entangled states. Below, we discuss the definitions of these, along with an introduction to positive maps and entanglement detection using some well-known positive maps.

\subsection{Positive maps in the realm of bipartite entanglement detection} \label{s2A}
We start by introducing the mathematical notion of bipartite entanglement and the role of positive maps in its detection. A bipartite state $\rho_{AB} \in \mathcal{D}(\mathbb{C}^{d} \otimes \mathbb{C}^{d})$ (with $\mathbb{C}^{d} \otimes \mathbb{C}^{d}$ representing the composite Hilbert space of systems $A$ and $B$) is said to be separable if and only if (iff) it can be written as 
\begin{equation}
    \rho_{AB} = \sum_{i} p_{i} \rho_{A}^{i} \otimes \rho_{B}^{i}     \label{separable}
\end{equation} 
where $\{p_{i}\}$ is a probability distribution and $\rho_{A}^i, \rho_{B}^i$ are valid density operators for subsystems $A$ and $B$ respectively. A state that does not satisfy Eq.~\eqref{separable} is said to be entangled. In the following, we present the formalism of entanglement detection using positive but not completely positive maps. A linear map $\Lambda :\mathcal{B}(\mathbb{C}^{d}) \rightarrow \mathcal{B}(\mathbb{C}^{d})$ is said to be:

(i) \textit{positive}: If $\Lambda (\rho) \ge 0$  $\forall \rho \ge 0$, with $\rho \in \mathcal{D}(\mathbb{C}^{d})$,

(ii) \textit{$k$-positive}: If $(\mathbb{1}_{k} \otimes \Lambda) (\rho_{AB}) \ge 0$ for some $k \in \mathbb{N}$, $\rho_{AB} \in \mathcal{D}(\mathbb{C}^{k} \otimes \mathbb{C}^{d})$  with $\mathbb{1}_k : \mathcal{B}(\mathbb{C}^{k}) \rightarrow \mathcal{B}(\mathbb{C}^{k})$ representing the $k$-dimensional identity map,

(iii) \textit{completely positive} (CP): If (ii) holds $\forall k \in \mathbb{N}$.

Although the definition suggests that determining complete positivity requires checking an infinite number of cases (one for each $k\in \mathbb{N}$), this task can be circumvented using the famous Choi-Jamiolkowski (CJ) isomorphism \cite{jamiolkowski1974effective, choi1975positive}. According to CJ isomorphism, a map is CP iff the corresponding Choi operator ($\mathcal{C}_{\Lambda}$) is positive, where $\mathcal{C}_{\Lambda}: = (\mathbb{1} \otimes \Lambda) \ket{\phi^+}\bra{\phi^+}$ with $\ket{\phi^+} = \frac{1}{\sqrt{d}} \sum_{i=0}^{d-1} \ket{ii} \in \mathcal{D}(\mathbb{C}^{d} \otimes \mathbb{C}^{d})$ being the maximally entangled state. A widely recognized example of a positive, yet not completely positive map is the transposition map, which plays a crucial role in entanglement detection. 

  The action of the transposition map on a bounded operator $X$ is defined as
    \begin{equation} 
        \Lambda_{\mathcal{T}}(X)= \begin{pmatrix}
x_{11} & x_{21}  \vspace{0.2cm}\\ 
x_{12} & x_{22}   
\end{pmatrix}  \label{transpositionmap}
    \end{equation} 
     where $\Lambda_{\mathcal{T}} : \mathcal{B}(\mathbb{C}^2) \rightarrow \mathcal{B}(\mathbb{C}^2)$ is the transposition map for qubits and $X = \begin{pmatrix}
x_{11} & x_{12}\\
x_{21} & x_{22}
\end{pmatrix} $. 
This map is optimal for entanglement detection in systems of dimension $\le 6$. According to the Peres-Horodecki criterion \cite{peres1996separability, horodecki2001separability}, a state $\rho$ with dim$(\rho) \le 6$ is separable iff $(\mathbb{1} \otimes \Lambda_{\mathcal{T}}) \rho \ge 0$. For dim $>6$, negativity under the transposition map is only a sufficient but not necessary criterion for entanglement since there also exist PPT (positive under partial transposition) entangled states. Therefore, by checking the negativity of the partially transposed state, one can only detect NPT entanglement.

Note that, beyond the transposition map, there exist several other positive but not completely positive maps that are also capable of detecting entangled states. A prominent example is the reduction map $\Lambda_{\mathcal{R}} : \mathcal{B}(\mathbb{C}^d) \rightarrow \mathcal{B}(\mathbb{C}^d)$, whose action is defined as follows:  \begin{equation}
       \Lambda_{\mathcal{R}} (X) = \Tr[X]. I_d -X   \label{reductionmap}
    \end{equation}  
    where $I_d$ is the $d$-dimensional identity matrix. 
 
A necessary condition for the separability of a quantum state $\rho$ via the reduction map is $(\mathbb{1} \otimes \Lambda_{\mathcal{R}})\rho \ge 0$ \cite{horodecki1999reduction, cerf1999reduction}, which is equivalent to the conditions $\rho_A \otimes I - \rho \ge 0$ and $I \otimes \rho_B - \rho \ge 0$, where $\rho_{A(B)} = \Tr_{B(A)} \rho$. These jointly constitute the reduction criterion for separability, and any violation certifies entanglement.

\subsection{Generation of positive maps from Lindblad structure}
In this subsection, we discuss how positive maps, such as transposition and reduction maps, can be systematically derived from the Lindblad structure of open quantum systems through suitable parameterization \cite{hall2014canonical,chimalgi2023detecting}.

The dynamics of open quantum systems evolving under system-environment interactions is usually governed by completely positive and trace-preserving (CPTP) maps. Such CPTP maps can be generated from Lindblad-type superoperators \cite{lindblad1976generators,alicki2007quantum}. Consider a system interacting with its surrounding environment under the influence of a specific interaction Hamiltonian. Assuming that the initial state of the combined system is in product form and by applying the stationary bath assumption along with the Born-Markov approximation, we can derive the master equation that governs the time evolution of the system's state. In certain cases, the Lindblad-type master equation can be obtained even without invoking the stationary bath assumption or the Born-Markov approximation. Hence, Lindblad-type structure plays a crucial role in the study of open quantum systems. If $\mathcal{L}_{\mathcal{x}}$ is the Lindblad generator corresponding to a positive map $\Lambda_{\mathcal{x}} : \mathcal{B}(\mathbb{C}^2) \rightarrow \mathcal{B}(\mathbb{C}^2)$, then

\begin{equation}
    \Lambda_{\mathcal{x}}(X) = (\mathbb{1}+\mathcal{L}_{\mathcal{x}})(X)  \label{mapfromLindblad}
\end{equation}
where $X = \begin{pmatrix}
x_{11} & x_{12}\\
x_{21} & x_{22}
\end{pmatrix}  \in \mathcal{B}(\mathbb{C}^2)$.
The action of a Lindblad generator on a bounded positive operator $X$ is 
\begin{equation}
   \mathcal{L}_{\mathcal{x}} (X) = \sum_{i} \gamma_{i} [\sigma_{i} X \sigma_{i}^{+} - \frac{1}{2}(\sigma_{i}^{+} \sigma_{i} X + X \sigma_{i}^{+} \sigma_{i})]  \label{Lindbladaction}
\end{equation}
where $\gamma_{i}$ are the Lindblad coefficients and $\sigma_{i}$ are the Pauli matrices.

To generate the transposition map defined in Eq.~\eqref{transpositionmap}, we take, $\gamma_1=\gamma_3=\frac{1}{2}$  and $\gamma_2=-\frac{1}{2}$, then Eq.~\eqref{mapfromLindblad} becomes 
\begin{equation} \label{lindbladmap_transpose}
\begin{split}
    \Lambda_{\mathcal{T}}(X) = X &+ \frac{1}{2} (\sigma_1 X {\sigma_1} - \frac{1}{2}\{{\sigma_1}  \sigma_1, X\} ) \\ &
   - \frac{1}{2} (\sigma_2 X {\sigma_2} - \frac{1}{2}\{{\sigma_2}  \sigma_2, X\} ) \\ &
   + \frac{1}{2} (\sigma_3 X {\sigma_3} - \frac{1}{2}\{{\sigma_3}  \sigma_3, X\} ) 
    \end{split}
\end{equation}
where $\{A,B\}:=AB+BA$. After simplification, Eq.~\eqref{lindbladmap_transpose} turns out to be
\begin{equation}\label{aciton}
     \Lambda_{\mathcal{T}}(X) = \begin{bmatrix} 
	x_{11}& x_{21} \\[0.2cm] 
	x_{12}& x_{22} \\
		\end{bmatrix}.
\end{equation}
Therefore, for specific values of the time-independent Lindblad coefficients $\gamma_i$, we can generate the transposition map $(\Lambda_{\mathcal{T}})$. 

Note that, we can also generate the reduction map defined in Eq.~\eqref{reductionmap} by choosing $\gamma_1=\gamma_2=\gamma_3=\frac{1}{2}$.\\

Moving beyond the bipartite regime, below we present the formalism of GME detection using these maps.

\subsection{Genuine multipartite entanglement detection}
In the multipartite case, there are various layers of separability. A state is said to be \textit{fully separable} if there is no entanglement across any of its bipartitions. Formally, an $N$-partite state $\rho_{sep} \in \mathcal{D}(\mathbb{C}^{d} \otimes \mathbb{C}^{d}\otimes....\otimes \mathbb{C}^{d})$ is called fully separable iff it can be written as 
\begin{equation}
    \rho_{sep} = \sum_{i} p_{i} \rho_{1}^{i} \otimes \rho_{2}^{i} \otimes....\rho_{N}^{i} \label{fullysep}
\end{equation}
where $\{p_i\}$ form a probability distribution, and each $\rho_{j}^{i}$ is a density matrix associated with subsystem $j$.
If a state cannot be written in the form of Eq.~\eqref{fullysep}, this indicates the signature of entanglement among some or all of its bipartitions. Depending on the number of layers across which entanglement is present, we have the notion of $k$-separability. A state $\rho_{2-sep}$ is said to be \textit{$2$-separable} or \textit{biseparable} if it can be written as 
\begin{equation}
    \rho_{2-sep} = \sum_{A}\sum_{i} p_{A}^{i} \rho_{A}^{i} \otimes \rho_{\bar{A}}^{i}  \label{2-sep}
\end{equation}
where $A (\bar{A})$ is a proper subset of the parties (complementary subset of the parties), i.e., $A \subset \{1,2,....,N\}$ and $\sum_{A}$ represents the sum over all bipartitions $A|\bar{A}$. A state that does not admit a decomposition of the form of Eq.~\eqref{2-sep} is said to be \textit{genuine $N$-partite entangled}. For example, consider the tripartite three-qutrit non-genuine entangled state
\begin{align}
\rho_{ABC}=\frac{1}{3}\left(\sum_{X\in\{A,B,C\}}\ket{2}\bra{2}_{X}\otimes\ket{\phi^+}_{\bar{X}}\bra{\phi^+}\right) \nonumber   
\end{align}
If all the three parties perform the measurement $\{\ket{0}\bra{0}+\ket{1}\bra{1},\ket{2}\bra{2}\}$ then, there is a $1/3$ chance that Alice and Bob observe the outcome $\ket{0}\bra{0}+\ket{1}\bra{1}$ and Charlie observes the outcome $\ket{2}\bra{2}$ and a $2/3$ chance of observing other outcomes in which case Alice and Bob discard their shared state and prepare the state $\ket{0}_A\bra{0}\otimes\ket{1}_B\bra{1}$. Thus, the effective state between Alice and Bob would be 
\begin{align}
\frac{1}{3}\ket{\phi^+}_{AB}\bra{\phi^+}+ \frac{2}{3}\left(  \ket{0}_A\bra{0}\otimes\ket{1}_B\bra{1}\right)
\end{align}
which is clearly entangled via rank deficit criteria \cite{guhne2009entanglement} (the state has matrix rank $2$, but the support of the state: Span$\{\ket{\phi^+},\ket{01}\}$ contains exactly one product state $\ket{01}$). This certifies that the original state $\rho_{ABC}$ is entangled in $A|BC$ and by symmetry in the other bipartitions as well. However the state $\rho_{ABC}$ is not GME as evident from the decomposition itself. Thus, certifying entanglement in all bipartitions (\(A|BC, B|AC, C|AB\)) does not certify GME simply because the state might not be separable in any bipartition but could lie in the convex hull of biseparable states. The presence of these layers makes the characterization of multipartite entanglement different and complex in comparison to the bipartite case. Moreover, multiple inequivalent classes of genuinely entangled states arise, rendering the notion of a `maximally entangled state' relative in the multipartite setting. In contrast, bipartite entanglement admits essentially a single entanglement class under stochastic local operations and classical communication, since all bipartite pure states with full Schmidt rank are equivalent to a maximally entangled state.

Despite these complexities, one of the promising tools to detect GME involves the use of positive maps, analogous to the bipartite case. We shall denote those positive maps that can be used for detecting GME to be \textit{GME maps} ($\Lambda_{\text{GME}}$). Any such GME map detecting GME should have the property 
\begin{equation}
    \Lambda_{\text{GME}}(\rho_{2-sep}) \ge 0 \; \forall \rho_{2-sep}. \label{GMEmaps}
\end{equation}
Violation of the above equation is a signature of GME. Below, we state the efficacy of the aforementioned maps in GME detection. 

Let $\Phi^{(N)}_{\mathcal{x}}$ be a Hermiticity preserving, linear map defined by 
\begin{equation}
    \Phi^{(N)}_{\mathcal{x}} (.) = \sum_{A}[\Lambda_{\mathcal{x}}^A \otimes \mathbb{1}_{\bar{A}} + c_{\mathcal{x}}^{(N)}. I_{2^N}. \Tr](.)  \label{map}
\end{equation}
where $\Lambda_{\mathcal{x}}^A : \mathcal{B}(\mathbb{C}^2) \rightarrow \mathcal{B}(\mathbb{C}^2)$ is a map having a Lindblad generator $\mathcal{L}_{\mathcal{x}}^A$, $I_{2^N}$ is the $2^N$-dimensional Identity matrix and $c_{\mathcal{x}}^{(N)}$ is chosen carefully such that Eq.~\eqref{GMEmaps} is satisfied. The minimum output eigen value of $\Lambda_{\mathcal{x}}$ is given by \cite{clivaz2017genuine}
\begin{equation}
    \nu(\Lambda_{\mathcal{x}}) = -\min_{\rho} \{EV_{\min}[(\mathbb{1} \otimes \Lambda_{\mathcal{x}}) \rho]\}  \label{mineigenvalue}
\end{equation}
where $EV_{\min}$ denotes the minimum eigen value of $(\mathbb{1} \otimes \Lambda_{\mathcal{x}}) \rho$. Furthermore, it has been shown that 
\begin{equation}
    c_{\mathcal{x}}^{(N)} \ge (2^{N-1}-2) \nu(\Lambda_{\mathcal{x}})  \label{valueofc}
\end{equation}
If we consider $\mathcal{x} = \mathcal{T}$, then $ \nu(\Lambda_{\mathcal{T}}) = \frac{1}{2}$ \cite{clivaz2017genuine}. For simplicity, throughout the paper we adopt the lower bound of $c_{\mathcal{T}}$ i.e. for the transposition map, we take $c_{\mathcal{T}}^{(N)}= \frac{2^{N-1}-2}{2} $.

However, a significant limitation of the positive map approach is its lack of direct physical implementability in experiments. In this context, we aim to detect (genuine) entanglement in an experimentally feasible way. With this aim, we discuss the partial transpose moment criterion below which provides a sufficient criterion for entanglement detection in an experiment-friendly way.

\subsection{Partial transpose moment criterion} 
As discussed previously, entanglement detection using the transposition map serves as an optimal criterion for bipartite entanglement detection of $\mathcal{D}(\mathbb{C}^2 \otimes \mathbb{C}^2), \mathcal{D}(\mathbb{C}^2 \otimes \mathbb{C}^3)$ and $\mathcal{D}(\mathbb{C}^3 \otimes \mathbb{C}^2)$ systems. However, this map being unphysical, can not be implemented directly in experiments. Thus, given an unknown quantum state, the task of entanglement detection reduces to performing quantum state tomography. However, this approach is resource-intensive, and its complexity grows exponentially with the number of subsystems and the dimensionality of each, making it impractical for multipartite or high-dimensional systems. In order to circumvent this, the idea of partial transpose moments (PT-moments) was introduced  \cite{calabrese2012entanglement}. The bipartite PT-moments are defined as 
\begin{equation}
    p_n = \Tr[{(\mathbb{1} \otimes \Lambda_{\mathcal{T}})\rho_{AB}}]^n \label{PTmoments}
\end{equation}
where $n \in \mathbb{N}$ represents the order of the moment and $\rho_{AB} \in \mathcal{D}(\mathbb{C}^d \otimes \mathbb{C}^d)$. For a normalized state, $p_1 = 1$, and $p_2$ is related to the purity of the state. Therefore, $p_3$ is the first non-trivial moment. Using the first three moments, a sufficient criterion for NPT entanglement was introduced, which is popularly known as the $p_3$-PPT criterion \cite{elben2020mixed}. For a PPT state, 
\begin{equation}
    p_{3} \ge p_{2}^2.  \label{p3PPT}
\end{equation}
Violation of Eq.~\eqref{p3PPT} indicates that the state is NPT, and hence entangled. For Werner states, the full PPT criterion and $p_3$-PPT criterion are equivalent. This can be extended to higher orders, and one can have a family of entanglement detection criteria using higher order moments with the $p_3$-PPT in the lowest order. In order to do so, the authors in \cite{yu2021optimal} introduce the notion of Hankel matrices $[H_{l}(\mathbf{p})]_{ij}$, where $i,j \in \{0, 1, ..., l\}$, $l \in \mathbb{N}$ and $\mathbf{p}=(p_1, p_2, ..., p_n)$. These are $(l+1) \times (l+1)$ matrices defined by 
\begin{equation}
    [H_{l}(\mathbf{p})]_{ij} = p_{i+j+1}  \label{Hankelmatrices}
\end{equation}
Therefore, the first and the second Hankel matrices are given by 
\begin{equation}
    H_1 (\mathbf{p}) = \begin{pmatrix}
p_1 & p_2   \vspace{0.2cm}\\ 
p_2 & p_3   \label{firstHankelmatrix}
\end{pmatrix} 
    \end{equation}
    and 
  \begin{equation}
    H_2 (\mathbf{p})= \begin{pmatrix}
p_1 & p_2 & p_3  \vspace{0.2cm}\\ 
p_2 & p_3 & p_4  \vspace{0.2cm}\\
p_3 & p_4 & p_5      \label{secondHankelmatrix}
\end{pmatrix}
    \end{equation} respectively. A necessary condition for separability using Hankel matrices is 
    \begin{equation}
        \det[H_{l}(\mathbf{p})] \ge 0 \hspace{1cm} \forall l \in \mathbb{N} . \label{Hankelmatrixcondition}
    \end{equation} 
    Note that $\det[H_1(\mathbf{p})] \ge 0$ is the $p_3$-PPT criterion (Eq.~\eqref{p3PPT}).

    This approach of using PT-moments for entanglement detection does not require any prior knowledge about the state, and hence is advantageous compared to witness-based detection schemes. Evaluation of such moments involves simple functionals that are easy to realize in experiments by a technique called shadow tomography \cite{huang2020predicting}. Unlike that in tomography, where the aim is to reconstruct the quantum state, this technique involves the evaluation of linear functions of the state. Since the quantities of interest usually involve linear functions of the state, e.g., entanglement witnesses, fidelity, etc.,  this serves as an efficient way to obtain such quantities. Further, this method is resource-effective since a polynomial number of state copies are sufficient to predict an exponential number of target functions \cite{aaronson2018shadow, aaronson2019gentle}. It may also be noted that evaluating all the higher order moments provides a necessary and sufficient criterion for NPT entanglement \cite{neven2021symmetry}. However, implementing all such moments is again experimentally challenging. Motivated by these, in the following section we propose a way to detect GME using truncated moments of the transposition map. (For a similar approach using moments of the reduction 
    map, see appendix \ref{A}).

\section{Moment-based genuine multipartite entanglement detection} \label{s3}
We start by defining the $n$-th order moments of the transposition map. 

\textbf{Definition 1:} If $\Phi^{(N)}_{\mathcal{x}}$ is a Hermiticity preserving, linear map given in Eq.~\eqref{map}, then we can define the $n$-th order moments of $\Phi^{(N)}_{\mathcal{x}}$ as 
\begin{equation}
    s_{n}^{(\mathcal{x})}= \Tr[\Phi^{(N)}_{\mathcal{x}} (\rho)]^n \label{moments}
\end{equation}
where $n \in \mathbb{N}$. For $\mathcal{x}=\mathcal{T}$, 
    \begin{equation} \label{transpositionmoments}
        s_n^{(\mathcal{T})} = \Tr[\Phi^{(N)}_{\mathcal{T}} (\rho)]^n
    \end{equation} represents the moments of the transposition map.
  
Using this definition, we now propose our criterion to detect GME.

\begin{theorem} \label{theorem1}
If a state $\rho_{2-\text{sep}} \in \mathcal{D}(\mathbb{C}^d \otimes \mathbb{C}^d \otimes ....\otimes \mathbb{C}^d )$ is biseparable, 
 \begin{equation}
        \det[H_{l}(\mathbf{s^{(\mathcal{T})}})] \ge 0  \label{multipartiteHankelmatrixcondition}
    \end{equation} 
    where $[H_{l}(\mathbf{s^{(\mathcal{T})}})]_{ij} = s_{i+j+1}^{(\mathcal{T})}$ for $i,j \in \{0,1,...,l\}$, $l \in \mathbb{N}$ and $\mathbf{s^{(\mathcal{T})}}=(s_1^{(\mathcal{T})}, s_2^{(\mathcal{T})},...,s_n^{(\mathcal{T})})$ is defined in Eq.~\eqref{moments}.
    \end{theorem}
    \proof Since $\Phi^{(N)}_{\mathcal{T}}$ is a Hermiticity preserving, linear map, for arbitrary $\rho \in \mathcal{D}(\mathbb{C}^d \otimes \mathbb{C}^d \otimes ....\otimes \mathbb{C}^d )$, $\Phi^{(N)}_{\mathcal{x}} (\rho)$ is Hermitian. Hence, $\Phi^{(N)}_{\mathcal{T}} (\rho)$ has a spectral decomposition, i.e., 
    \begin{equation}
        \Phi^{(N)}_{\mathcal{T}} (\rho) = \sum_{i=1}^{d^N} \lambda_{i} \ket{i}\bra{i}  \label{spectral}
    \end{equation}
    Therefore, from Eq.~\eqref{moments}, $s_{n}^{(\mathcal{T})} = \sum_{i=1}^{d^N} (\lambda_{i})^n$. Note that $s_1^{(\mathcal{T})}=1$ for any positive map $\Lambda_{\mathcal{T}}$. The Hankel matrices defined in Eq.~\eqref{Hankelmatrices} admit a decomposition of the form \cite{heinig1984algebraic, tyrtyshnikov1994bad}
    \begin{equation}
        H_{l}(\mathbf{s^{(\mathcal{T})}}) = V_{l} D V_{l}^{\mathcal{T}}  \label{Hankelmatrixdecomposition}
    \end{equation}
    where $\mathcal{T}$ denotes transpose in the computational basis, $D$ is a diagonal matrix given by $D = \text{diag} \{\lambda_1, \lambda_2,...,\lambda_{d^N}\}$ and 
    \begin{equation}
        V_{l} = \begin{bmatrix} 
    1 & 1 & \dots & 1\\
    \lambda_{1} & \lambda_{2} & \dots & \lambda_{d^N}\\
    \vdots & \vdots & \ddots & \vdots\\
    \lambda_{1}^{l} & \lambda_{2}^{l} & \dots & \lambda_{d^N}^{l} 
    \end{bmatrix}.  \label{Vl}
    \end{equation}
    If $\rho$ is biseparable, i.e., $\rho \equiv \rho_{2-\text{sep}}$, then $\Phi^{(N)}_{\mathcal{T}} (\rho_{2-\text{sep}}) \ge 0$ for any $\Lambda_{\mathcal{T}}$ and by choosing a suitable value of $c_{\mathcal{T}}$ given by Eq.~\eqref{valueofc}. Therefore, $\lambda_{i} \ge 0$  $\forall i \in \{1, 2, ..., d^N\}$. \\
    
    Let $\mathbf{y}=(y_1, y_2,...,y_{l+1})$ be an arbitrary vector belonging to $\mathbb{R}^{l+1}$. Using Eq.~\eqref{Hankelmatrixdecomposition}, $\mathbf{y} H_{l}(\mathbf{s^{(\mathcal{T})}}) \mathbf{y}^{\mathcal{T}} = \mathbf{z} D \mathbf{z}^{\mathcal{T}}$  where $\mathbf{z} = \mathbf{y} V_{l} = (z_1, z_2, ..., z_{d^N})$ and $z_{i} = y_1 + \sum_{j=1}^{l} (\lambda_{i})^{j}$  $x_{j+1}$ for $i=1,2, ...., d^{N}$.
    $$\mathbf{y} H_{l}(\mathbf{s^{(\mathcal{T})}}) \mathbf{y}^{\mathcal{T}} = \mathbf{z} D \mathbf{z}^{\mathcal{T}} = \sum_{i=1}^{d^N} \lambda_{i} z_{i}^2 \ge 0 \hspace{0.2cm} \forall \mathbf{y} \in \mathbb{R}^{l+1}.$$ This implies that $H_l(\mathbf{s^{(\mathcal{T})}}) \ge 0$ and hence, 
    $\det[H_{l}(\mathbf{s^{(\mathcal{T})}})] \ge 0$. This completes the proof.   \qed  
    
    Now from the contrapositive statement, violation of Eq.~\eqref{multipartiteHankelmatrixcondition} is sufficient to conclude that the multipartite state is genuinely entangled. We present some examples in the tripartite scenario below to show the effectiveness of our criterion. Our approach is general enough and can be readily extended to an arbitrary $N$-partite system in a similar fashion.
    
\subsection{Examples for $N=3$}
Following Eq.~\eqref{map}, next we will use the transposition-based GME map to detect genuine entanglement in tripartite systems, which is defined as 
\begin{equation}
\begin{split}
    \Phi^{(3)}_{\mathcal{T}}(*) = & (\mathbb{1} \otimes \mathbb{1} \otimes \Lambda_{\mathcal{T}} + \mathbb{1} \otimes \Lambda_{\mathcal{T}} \otimes \mathbb{1} + \Lambda_{\mathcal{T}} \otimes \mathbb{1} \otimes \mathbb{1} \\ & + I_8. \Tr) (*) 
    \end{split}  \label{tripartitemap}
\end{equation} 
where $I_8$ is the $8\times8$ identity matrix and $c^{(3)}_{\mathcal{T}}$ is taken to be $1$. In tripartite systems, there are two inequivalent classes of genuine tripartite entanglement, namely the $W$ and $GHZ$ states \cite{dur2000three}. Below, we apply our moment-based approach to detect the genuine entanglement of these two classes.

\begin{example} \label{example1}
Detection of $GHZ$ state using moments of modified transposition map:\\
Consider the $GHZ$ state, where $\ket{GHZ ^{(3)}} \in \mathcal{D}(\mathbb{C}^{2} \otimes \mathbb{C}^{2} \otimes \mathbb{C}^{2})$ is defined as
\begin{equation}
\ket{GHZ ^{(3)}}= \frac{\ket{000}+\ket{111}}{\sqrt{2}} . \label{GHZ}
\end{equation}

To detect the GME of the $GHZ$ state, we introduce a modified linear map (the modified transposition map), defined as
 \begin{equation}	 \label{modified}
 \begin{split}
      \tilde{ \Phi}^{(3)}_{\mathcal{T}}(*) & = [(\sigma_x \circ \Lambda_{\mathcal{T}})\otimes \mathbb{I}_{2} \otimes \mathbb{I}_{3}+ \mathbb{I}_{1} \otimes (\sigma_x \circ \Lambda_{\mathcal{T}})  \otimes \mathbb{I}_{3} \\
      & + \mathbb{I}_{1} \otimes \mathbb{I}_{2} \otimes (\sigma_x \circ \Lambda_{\mathcal{T}}) + I_8.\Tr](*)
       \end{split}
	 \end{equation} 
  where $\sigma_x \circ\Lambda_{\mathcal{T}}$ represents the composition of the transposition map followed by a unitary operation $\sigma_x$. Since local unitary operations such as $\sigma_x$ can not enhance entanglement, therefore the modified map $\tilde{\Phi}_{\mathcal{T}}^{(3)}$ remains positive on all biseparable states. Moreover the modified map given by Eq.~\eqref{modified} acts as a GME map, since $\tilde{\Phi}_{\mathcal{T}}^{(3)}(\ket{GHZ}\bra{GHZ})$ gives us a negative eigenvalue and is positive on all biseparable states \cite{clivaz2017genuine}.

Now, to detect the $GHZ$ state using moment-based criteria, we define the modified transposition moments as 
\begin{equation} \label{modifiedmoment}
      \tilde{s}_n^{(\mathcal{T})} = \Tr[\tilde{\Phi}^{(N)}_{\mathcal{T}} (\rho)]^n
\end{equation}
It is important to note that the proof of Theorem \ref{theorem1} holds true in this context, as the structure of the argument is unaffected by the application of the local unitary operation $\sigma_x$. Consequently, the condition stated in Theorem \ref{theorem1} remains valid, since local unitaries cannot increase entanglement.

 Utilizing moments up to the third order, we obtain $\det[H_1(\mathbf{\tilde{s}^{(\mathcal{T})}})] < 0$. These results indicate that the modified transposition moments defined in Eq.~\eqref{modifiedmoment} can detect the genuine entanglement of the $GHZ$ state.
\end{example}

\begin{example}  \label{example2}
Detection of $W$ state using moments of transposition map:\\ 
Consider the $W$ state represented by $\ket{W ^{(3)}} \in \mathcal{D}(\mathbb{C}^{2} \otimes \mathbb{C}^{2} \otimes \mathbb{C}^{2})$ where
\begin{equation}
    \ket{W ^{(3)}}=\frac{\ket{001}+\ket{010}+\ket{100}}{\sqrt{3}}.
\end{equation}
The tripartite map used to detect this state is given by Eq.~\eqref{tripartitemap} where $\Lambda_{\mathcal{T}}$ is the qubit transposition map defined in Eq.~\eqref{transpositionmap} and $c_{\mathcal{T}}^{(3)} = 1.$

Note that $\Phi^{(3)}_{\mathcal{T}}(\ket{W ^{(3)}}\bra{W ^{(3)}})$ has a negative eigenvalue and hence, the state is genuinely entangled. However, the first Hankel matrix condition cannot detect the genuine entanglement of this state. Nevertheless, using the second Hankel matrix criterion, we obtain $\det[H_2(\mathbf{s}^{(\mathcal{T})})] < 0$. Hence, the genuine entanglement of $W$ state can be detected by the moments of the transposition map.\\
\end{example}

\begin{figure}[ht]
\includegraphics[width=.45\textwidth]{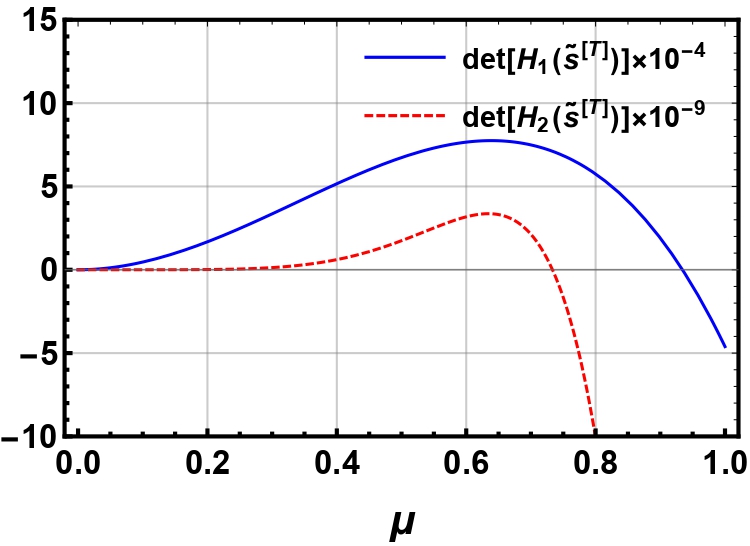} 
\caption{Detection of genuine entanglement of tripartite noisy $GHZ$ state using moments of modified transposition map.}\label{fig1}
\centering
\end{figure}

\begin{example} \label{example3}
Detection of noisy GHZ state using moments of modified transposition map: \\
Consider the noisy $GHZ$ state represented by 
\begin{equation}
    \ket{GHZ ^{(3)}}_\mu\bra{GHZ ^{(3)}} = \mu \ket{GHZ ^{(3)}}\bra{GHZ ^{(3)}} + \frac{(1-\mu)}{8} I_{8} \label{noisyGHZ}
\end{equation}
where $\mu \in [0, 1]$ is the noise parameter. Note that, the GME map $\tilde{\Phi}^{(3)}_{\mathcal{T}}$ detects the above tripartite noisy $GHZ$ state for $\mu > 0.733$ \cite{clivaz2017genuine}. 

Applying our proposed criterion defined in Eq.~\eqref{multipartiteHankelmatrixcondition}, and using the modified transposition moments, we observe that the first Hankel matrix becomes negative for $\mu > 0.934$. Therefore, for $0.733 < \mu \le 0.934$, the state is genuinely entangled, but the first Hankel matrix criterion is unable to detect the entire range. However, the second Hankel matrix detects the GME of this state for $\mu > 0.733$. This suggests that the map-based criteria introduced in \cite{clivaz2017genuine} and our modified transposition moment-based criteria are equivalent for the detection of noisy $GHZ$ state for moments up to \textit{fifth} order. This is shown in Fig. \ref{fig1}.
\end{example}

\begin{figure}[ht]
\includegraphics[width=0.45\textwidth]{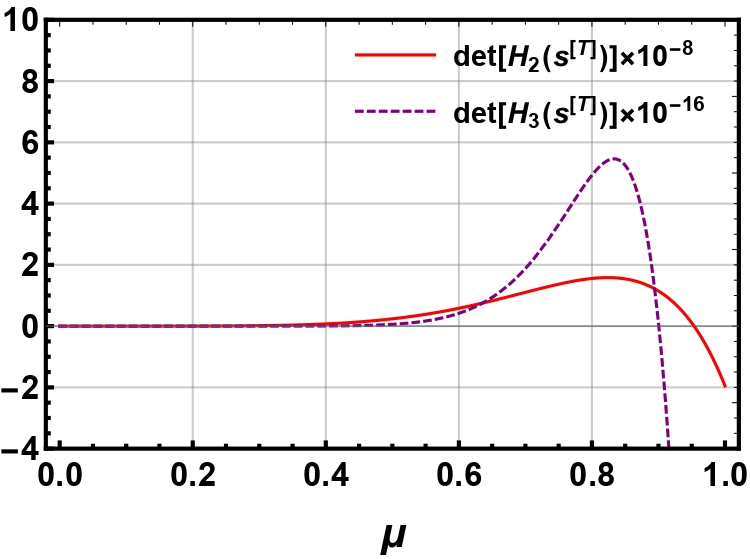}
\caption{Detection of genuine entanglement of tripartite noisy $W$ state using moments of transposition map.} 
\label{fig2}
\centering
\end{figure}

\begin{example}  \label{example4}
Detection of noisy $W$ state using moments of transposition map: \\
Consider the noisy $W$ state defined as
\begin{equation}
    \ket{W ^{(3)}}_\mu\bra{W ^{(3)}} = \mu \ket{W ^{(3)}}\bra{W ^{(3)}} + \frac{(1-\mu)}{8} I_{8} \label{noisyW}
\end{equation} where $\mu \in [0,1]$. The map given in Eq.~\eqref{tripartitemap} can detect the genuine entanglement of this state for $\mu > 0.899$. In contrast, the condition $\det[H_2(\mathbf{s}^{(\mathcal{T})})] < 0$ is satisfied only for $\mu > 0.953$. Therefore, in the range $0.899 < \mu \leq 0.953$, the entanglement remains undetected by the second Hankel matrix criterion. However, using moments up to the seventh order, the entire range is detected. So, the map-based criterion and the moment-based criterion are equivalent for moments up to \textit{seventh} order. Fig.\ref{fig2} illustrates the violation of the second and third Hankel matrix conditions as a function of the noise parameter $\mu$. 
\end{example}

\begin{example}  \label{example5}
Detection of convex mixture of $GHZ$ and $W$ state using moments of modified transposition map:\\
Let $\ket{\psi^{(3)}}\bra{\psi^{(3)}}$ be the tripartite state $\in \mathcal{D}(\mathbb{C}^{2} \otimes \mathbb{C}^{2} \otimes \mathbb{C}^{2})$, represented by
\begin{equation}
\begin{split}
    \ket{\psi^{(3)}}\bra{\psi^{(3)}} =& \mu \ket{GHZ^{(3)}}\bra{GHZ^{(3)}} \\ & + (1-\mu) \ket{W^{(3)}}\bra{W^{(3)}}.
    \end{split}
\end{equation}  \label{convexmixture}
 The GME map $\tilde{\Phi}^{(3)}_{\mathcal{T}}$ can detect $\ket{\psi^{(3)}}\bra{\psi^{(3)}}$ for the region $\mu > 0.746$.

Now, if we apply our proposed criterion defined in Eq.~\eqref{multipartiteHankelmatrixcondition}, based on the modified transpose moments, we find that the above mentioned state is detected by the first and second Hankel matrix conditions for $\mu > 0.945 $ and $\mu > 0.755$, respectively. However, using the third Hankel matrix criterion, the entire range is detected. In other words, the map-based criterion and our moment-based criterion are equivalent for moments up to \textit{seventh} order. This is shown in Figure \ref{fig3}.
\begin{figure}[ht]
    \centering
    \includegraphics[width=0.45\textwidth]{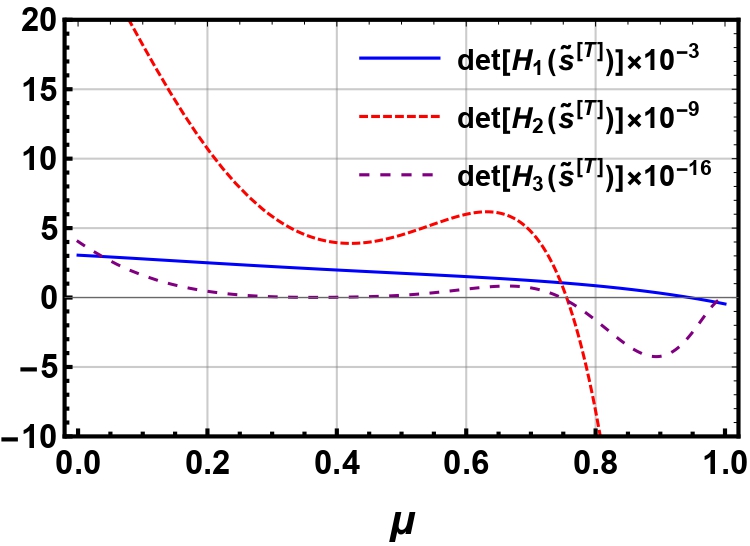}
    \caption{Detection of genuine entanglement of a convex mixture of $GHZ$ and $W$ state using moments of modified transposition map.}
    \label{fig3}
\end{figure}
\end{example}
Our findings show that as the order of moments increases, the detection criteria become more stringent, potentially leading to tighter bounds. Moreover, some of the states presented above can also be detected using the moments of other positive maps, e.g., the reduction map. This is discussed in the appendix \ref{A}.

\subsection{Example for $N=4$} To show the efficacy of our method for multipartite systems, we provide an example for the detection of a GME state beyond the tripartite case. 

Following the construction in Eq. \eqref{map}, we use the modified transposition map with $c_{\mathcal{T}}^{(4)} = 3$ as given below:
\begin{equation}	 \label{modifiedquadripartitemap}
 \begin{split}
      \tilde{ \Phi}^{(4)}_{\mathcal{T}}(*) & = [(\sigma_x \circ \Lambda_{\mathcal{T}})\otimes \mathbb{I}_{2} \otimes \mathbb{I}_{3} \otimes \mathbb{I}_{4} \\ &+ \mathbb{I}_{1} \otimes (\sigma_x \circ \Lambda_{\mathcal{T}})  \otimes \mathbb{I}_{3} \otimes \mathbb{I}_{4}\\
      & + \mathbb{I}_{1} \otimes \mathbb{I}_{2} \otimes (\sigma_x \circ \Lambda_{\mathcal{T}}) \otimes  \mathbb{I}_{4} \\ &+ \mathbb{I}_{1} \otimes \mathbb{I}_{2} \otimes \mathbb{I}_{3} \otimes (\sigma_x \circ \Lambda_{\mathcal{T}}) \\ & + (\sigma_x \circ \Lambda_{\mathcal{T}}) \otimes (\sigma_x \circ \Lambda_{\mathcal{T}}) \otimes \mathbb{I}_{3} \otimes \mathbb{I}_{4} \\ & + (\sigma_x \circ \Lambda_{\mathcal{T}}) \otimes \mathbb{I}_{2} \otimes (\sigma_x \circ \Lambda_{\mathcal{T}})  \otimes \mathbb{I}_{4} \\ & + (\sigma_x \circ \Lambda_{\mathcal{T}}) \otimes \mathbb{I}_{2} \otimes \mathbb{I}_{3} \otimes (\sigma_x \circ \Lambda_{\mathcal{T}}) + 3 I_{16}.\Tr](*)
       \end{split}
	 \end{equation} 
     The moments corresponding to this map are given by Eq. \eqref{modifiedmoment} with $N=4$. We now pick some explicit examples from the quadripartite case.
\begin{example}
    Detection of $GHZ$ state using moments of the modified transposition map:\\
    Consider the $4$-qubit $GHZ$ state given by
    \begin{equation}
\ket{GHZ ^{(4)}}= \frac{\ket{0000}+\ket{1111}}{2} . \label{GHZ1}
\end{equation}
For this state, we obtain that $\det[H_1(\mathbf{\tilde{s}^{(\mathcal{T})}})] > 0$ and $\det[H_2(\mathbf{\tilde{s}^{(\mathcal{T})}})] < 0$. This shows that the fifth order moments corresponding to the modified transposition map is able to detect the $4$-qubit $GHZ$ state.
\end{example} 

\begin{example}
    Detection of the noisy $GHZ$ state using moments of the modified transposition map:\\
The $4$- qubit noisy $GHZ$ state is given by 
\begin{equation}
    \ket{GHZ ^{(4)}}_\mu\bra{GHZ ^{(4)}} = \mu \ket{GHZ ^{(4)}}\bra{GHZ ^{(4)}} + \frac{(1-\mu)}{16} I_{16} \label{noisyGHZ1}
\end{equation}
The map given by Eq. \eqref{modifiedquadripartitemap} detects this noisy $GHZ$ state for $\mu > 0.873$. Using moments of the modified transposition map, we obtain $\det[H_1(\mathbf{\tilde{s}^{(\mathcal{T})}})] > 0$, indicating that the first Hankel matrix criterion is not able to detect this state. However, the second Hankel matrix criterion successfully detects the state for the entire range of the noise parameter, i.e., $\det[H_2(\mathbf{\tilde{s}^{(\mathcal{T})}})] < 0$ for $\mu > 0.873$. This is illustrated in Figure \ref{fig6}.
\begin{figure}[ht]
    \centering
    \includegraphics[width=0.45\textwidth]{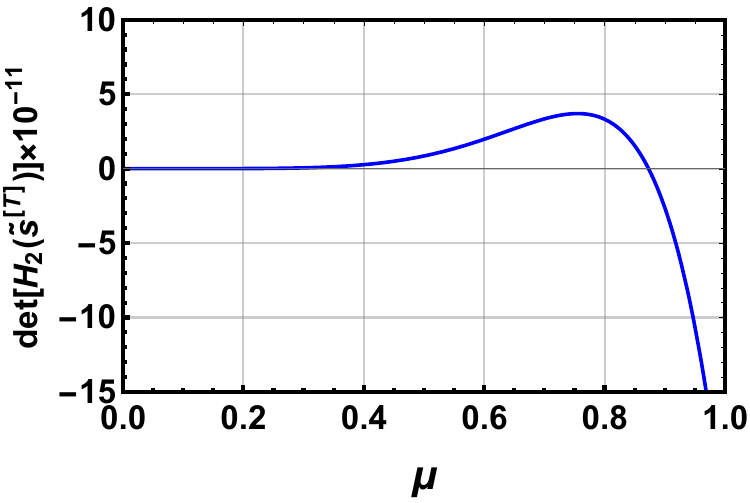}
    \caption{Detection of genuine entanglement of $4$-qubit noisy $GHZ$ state using moments of modified transposition map.}
    \label{fig6}
\end{figure}
\end{example}
It should be noted that, although we have presented explicit examples for the $3$-qubit and $4$-qubit cases, our method is applicable for the detection of $n$-qubit states, as well as to systems with arbitrary local dimensions. In such scenarios, one simply needs to select the appropriate positive map(s) tailored to the detection of the corresponding states.

\section{Proposal for experimental realization of the moments} \label{s4}
In this section, we present an experimentally realizable proposal to obtain the truncated moments of the transposition map used in our GME detection scheme. Calculating the $n$th-order moments involves preparing at most $n$ copies of the quantum state and finding the expectation value of suitably chosen operator(s) on that state \cite{PhysRevLett.121.150503,cieslinski2024analysing}. The specific form of the operator(s) depends on both the positive map employed and the order of the moment. Our moments are experimentally computable by finding the expectation values of these operators. Below, we provide the steps to realize the second, third, and $n$th order moments for the transposition map. For convenience, we start with the simplest case of a tripartite system with local dimension two, i.e., $N=3$, and $d=2$. This method can be extended for the realization of moments corresponding to various other maps in a similar fashion.

\begin{itemize}
   \item \textbf {Realization of the second order moment :}\\
     Here, we provide a prescription to find the second-order moment for the transposition map.
     Consider a tripartite state $\rho \in \mathcal{D}(\mathbb{C}^{2} \otimes \mathbb{C}^{2} \otimes \mathbb{C}^{2})$ and the tripartite map given by 
     Eq.~\eqref{tripartitemap}.
     The second order moment corresponding to this map is defined as 
     \begin{equation}
     \begin{split}
         s_2^{(\mathcal{T})} & = \Tr[\Phi_{\mathcal{T}}^{(3)} (\rho)]^2  \\ & =\Tr[\sum_{i=1}^{3}\Phi_{i} (\rho) + \Phi_{4} (\rho)]^2
         \end{split}    \label{modifiedsecondmoments}
   \end{equation} where $\Phi_{i} (\rho) = \otimes_{k=1}^{3} M_{k}^{(i)}$ for $i \in \{1,2,3\}$, with 
   \begin{align*}
M_{k}^{(i)} = 
\left\{
    \begin {aligned}
         & \Lambda_{\mathcal{T}} \quad & \text{for} \hspace{1mm} i=k \\
         & \mathbb{1} \quad & \text{for} \hspace{1mm} i \neq k                  
    \end{aligned}
\right.
\end{align*} and $\Phi_{4} (\rho) = I_8$. Here, $k=1 (2,3)$ represents the indices of three parties sharing the tripartite state, say, Alice (Bob, Charlie). Eq.~\eqref{modifiedsecondmoments} follows from the linearity of the map.
     Here, we provide a method to realize the individual terms appearing in the expression of $s_2^{(\mathcal{T})}$. The detailed calculation for the individual terms are given in the appendix \ref{B1}.
    Following Eq.~\eqref{modifiedsecondmoments}, the second order moments can be written as:
     \begin{equation}
     \begin{split}
         s_2^{(\mathcal{T})}  &=   \text{Tr}[\sum_{i=1}^{4} (\Phi_{i}(\rho))^2 + \sum_{\substack{i,j=1 \\ i \neq j}}^3 (\Phi_{i} (\rho) \Phi_{j}(\rho)) \\ &  ~~~~ +  2 \sum _{i=1} ^{3} (\Phi_{i} (\rho) \Phi_{4}(\rho))] \\ & = 8 +  \sum_{\substack{i,j=1 \\ (i,j) \neq (4,4)}}^4 \Tr[\Phi_i(\rho) \Phi_j (\rho)] 
         \end{split}  \label{secondmomentexperimental}
     \end{equation}
     Note that since the transposition is a trace-preserving (TP) map, $\Tr[\sum_{i=1}^3 \Phi_{i}(\rho) \Phi_4 (\rho)] = \Tr[\sum_{i=1}^3 \Phi_4 (\rho) \Phi_{i}(\rho)] = 3$. Therefore, Eq.~\eqref{secondmomentexperimental} reduces to 
    \begin{equation}
       s_2^{(\mathcal{T})} = 14 + 2 \sum_{\substack{i,j=1 \\ i < j}}^3 \Tr[\Phi_i(\rho) \Phi_{j} (\rho)] + \sum_{i=1}^{3} \Tr[(\Phi_i (\rho))^2]  .\label{secondordermomentsfinal}
    \end{equation}
   Hence, the expression for the second order moment consists of two terms.  Below, we present a way to realize these two terms in an experimental setup.
    \begin{enumerate}

    \item Realization of $\Tr[(\Phi_{i}(\rho))^2],$ for $ i = \{1,2,3\}$ :\\
    Note that
     \begin{equation}
         \Tr[(\Phi_{i}(\rho))^2] = \Tr[(\rho_{A B C} \otimes \rho_{A^{'} B^{'} C^{'}}) (X_A \otimes X_B \otimes X_C)]
     \end{equation}
    where $X_u = SWAP_{u u^{'}}$ for $u \in \{A,B,C\}$, is a Hermitian operator given by  
     \begin{equation}
     \begin{split}
     SWAP_{u u^{'}} = \begin{bmatrix}
     1 & 0 & 0 & 0 \\
     0 & 0 & 1 & 0  \\
     0 & 1 & 0 & 0  \\
     0 & 0 & 0 & 1 \\ 
     \end{bmatrix}_{u u^{'}}\label{SWAP}
 \end{split}
 \end{equation} 
 in the computational basis. Two copies of the unknown state are given. These terms are realized by finding the expectation value of $\otimes_u  X_u$ on two copies of the state. 
 \item Realization of $\Tr[\Phi_{i}(\rho) \Phi_{j}(\rho)], $ with $i,j = \{1,2,3\}, \text{ and } i \neq j$ :
 \begin{equation}
         \Tr[\Phi_{i}(\rho) \Phi_{j}(\rho)] = \Tr[(\rho_{A B C} \otimes \rho_{A^{'} B^{'} C^{'}}) (\otimes _{k}Y_{k})] \nonumber
     \end{equation} 
     where $k, k^{'}=1 (2,3)$ represents the indices of Alice's (Bob's, Charlie's) first and second particles respectively, and 
     \begin{equation}
     Y_{k} = 
\left\{
    \begin {aligned}
         & SWAP \quad & \text{for} \hspace{1mm} k \neq i,j \\
         &\hat{\phi} = 2 \ket{\phi^+}\bra{\phi^+} \quad & \text{for} \hspace{1mm} k =i \hspace{1mm}\text{or} \hspace{1mm} k=j                 
    \end{aligned}
\right.  \label{Y}
\end{equation}
where $\ket{\phi^+} = \frac{1}{\sqrt{2}} \sum_{i=0}^{1} \ket{ii}$. Note that Alice, Bob, and Charlie can realize these terms by finding the expectation value of $\otimes_k Y_k$ on the given two copies of the unknown state.

Since all the terms can be realized in real experiments by finding the expectation values of some local operators on the unknown state, the second order moment can be evaluated.

 \end{enumerate}
     \item \textbf {Realization of the third order moment :}\\ 
     One can realize the third order moment corresponding to the transposition map in an experimental setup by following the prescription given below. For a tripartite state $\rho$, the third order moments corresponding to the map defined in Eq.~\eqref{tripartitemap} can be expressed as follows
     \begin{equation}
     \begin{split}
         s_3^{(\mathcal{T})} = & \Tr[\Phi_{\mathcal{T}}^{(3)} (\rho)]^3  \\ & =\Tr[\sum_{i=1}^{3}\Phi_{i} (\rho) + \Phi_{4} (\rho)]^3
         \end{split}    \label{modifiedthirdmoments}
   \end{equation} where $\Phi_{i} (\rho) = \otimes_{k=1}^{3} M_{k}^{(i)}$ for $i \in \{1,2,3\}$,  
   \begin{align*}
M_{k}^{(i)} = 
\left\{
    \begin {aligned}
         & \Lambda_{\mathcal{T}} \quad & \text{for} \hspace{1mm} i=k \\
         & \mathbb{1} \quad & \text{for} \hspace{1mm} i \neq k                  
    \end{aligned}
\right.
\end{align*} and $\Phi_{4} (\rho) = I_8$. $k=1 (2,3)$ represents the indices of Alice (Bob, Charlie). Eq.~\eqref{modifiedthirdmoments} reduces to 
\begin{equation}
     \begin{split}
         s_3^{(\mathcal{T})}  = &  \text{Tr}[\sum_{i=1}^{4} (\Phi_{i}(\rho))^3] + 3\sum_{\substack{i,j=1 \\ i \neq j}}^4 (\Phi_{i} (\rho))^2 \Phi_{j}(\rho) \\ &   + \sum_{\substack{i,j,k=1 \\ i \neq j \neq k}}^4 \Phi_i (\rho) \Phi_j (\rho) \Phi_k (\rho)]\\ & = 8 +  \sum_{i=1}^3 \Tr[(\Phi_i(\rho))^3 +  3(\Phi_i (\rho))^2 \Phi_4 (\rho)] \\ & + 3\sum_{i,j=1}^3 \Tr[(\Phi_i(\rho))^2 \Phi_j (\rho)]  \\  & + \sum_{\substack{i,j,k=1 \\ i \neq j \neq k}}^3 \Tr[\Phi_i (\rho) \Phi_j (\rho) \Phi_k (\rho)]  \\ & + 3\sum_{\substack{i,j \\ i \neq j}}^3 \Tr[ \Phi_i (\rho) \Phi_j (\rho)\Phi_4 (\rho)] \\ & + 3\sum_{i=1}^3 \Tr[(\Phi_4 (\rho))^2\Phi_i (\rho)]
         \end{split} \label{thirdmomentsexperimental}
     \end{equation} 
Among these terms, $\sum_{i=1}^3 \Tr[(\Phi_i (\rho))^2 \Phi_4 (\rho)]$ and $\sum_{\substack{i,j \\ i \neq j}}^3 \Tr[ \Phi_i (\rho) \Phi_j (\rho)\Phi_4 (\rho)]$ are equivalent to the second order moment terms (since $\Phi_4 (\rho) = I_8$). Hence, these terms can be realized by the prescription described above for the second order moments. Moreover, since the transposition is a TP map, $\Tr[(\Phi_4 (\rho))^2\Phi_i (\rho) ] = 1$ $\forall i =1,2,3$. Therefore, Eq.~\eqref{thirdmomentsexperimental} becomes 
\begin{equation}
\begin{split}
    s_3^{(\mathcal{T})}  = & 17 + \sum_{i=1}^3 \Tr[(\Phi_i (\rho))^3] + 3\sum_{\substack{i,j=1 \\ i \neq j}}^3 \Tr[(\Phi_{i} (\rho))^2 \Phi_{j}(\rho)] \\ & + \sum_{\substack{i,j,k=1 \\ i \neq j \neq k}}^3 \Tr[\Phi_i (\rho) \Phi_j (\rho) \Phi_k (\rho)]\\ & + 3\underbrace{\sum_{\substack{i,j=1 \\ i \neq j}}^3 \Tr[\Phi_i (\rho) \Phi_j (\rho)]+ 3\sum_{i=1}^3 \Tr[(\Phi_i (\rho))^2]}_{\text{second order moment terms}}.
\end{split}   \label{thirdordermomentsfinal}
     \end{equation} 
     As an example, we discuss the realization of $\sum_{i=1}^3 \Tr[(\Phi_i (\rho))^3]$ below. The expressions for the other terms are given in the appendix \ref{B2}.
    \begin{enumerate}

    \item Realization of $\Tr[(\Phi_{i}(\rho))^3], \text{ for } i = \{1,2,3\}$ :\\
    Note that
    \end{enumerate}
     \begin{equation}
         \Tr[(\Phi_{i}(\rho))^3] = \Tr[(\rho_{A_1 B_1 C_1} \otimes \rho_{A_{2} B_{2} C_{2}} \otimes \rho_{A_{3} B_{3} C_{3}}) (\otimes_k Z_k)]
     \end{equation}
    where \begin{equation}
     Z_{k} = 
\left\{
    \begin {aligned}
         & SWAP_{k_1 k_2} SWAP_{k_2 k_3}\quad & \text{for} \hspace{1mm} k = i \\
         &SWAP_{k_1 k_3} SWAP_{k_2 k_3} \quad & \text{for} \hspace{1mm} k \neq i                
    \end{aligned}  \label{hermitianoperator}
\right.
\end{equation}
$k=1 (2,3)$ represents the parties Alice (Bob, Charlie). $SWAP_{k_1 k_2} SWAP_{k_2 k_3}$ indicates the $k$th party swapping her (his) second and third particles, followed by a swapping of the first and second particles. This is the backward $SWAP$ operator. Likewise, $SWAP_{k_1 k_3} SWAP_{k_2 k_3}$ is the forward $SWAP$ operator. In order to realize the above term, three copies of the unknown state are required. The expectation value of $\otimes_k Z_k$ is calculated to obtain the desired value. Note that for the second and third order moments corresponding to the transposition map, the relevant local operators are just the $SWAP$ and $\hat{\phi}$, but for other maps (e.g., the reduction map), one needs to choose the operator suitably. Nevertheless, for any positive map, there exists an appropriate operator whose expectation values give the associated moments \cite{cieslinski2024analysing}.  

The realization of the remaining terms can be done by finding the expectation values of the suitable operators on at most three copies of the unknown state. Hence, the third order moment can be evaluated. Here, it may be noted that for the case of detection of bipartite photon polarization entanglement, 
the scheme for experimental realization of the second, third, and fourth order moments has been presented earlier in \cite{PhysRevA.91.032315}.

 \item \textbf {Realization of $n$th order moments:}\\ 

 The higher order moments can be calculated by extending the proposed formalism.  Consider a state $\rho \in \mathcal{D}(\mathbb{C}^2 \otimes \mathbb{C}^2 \otimes \mathbb{C}^2)$. The $n$th order moments corresponding to the positive map $\Lambda_{\mathcal{x}}$ are defined as
\begin{equation}
    s_{n}^{(\mathcal{x})}= \Tr[\Phi^{(3)}_{\mathcal{x}} (\rho)]^n \label{moments1}
\end{equation} where $\Phi^{(3)}_{\mathcal{x}}(\rho)= \Phi_1 (\rho)+\Phi_2(\rho)+\Phi_{3}(\rho)+\Phi_4(\rho)$, and $\Phi_{i} = \otimes_{l=1}^{3} M_{l}^{(i)}$ for $i \in \{1,2,3\}$, with 
\begin{align*}
M_{l}^{(i)} = 
\left\{
    \begin {aligned}
         & \Lambda_{\mathcal{x}} \quad & \text{for} \hspace{1mm} i=l \\
         & \mathbb{I} \quad & \text{for} \hspace{1mm} i \neq l                  
    \end{aligned}
\right.
\end{align*} and $\Phi_{4} (\rho) = c_{\mathcal{x}}^{(3)} I_{8}$. Since $N=3$ throughout, we use $\Phi^{(3)}_{\mathcal{x}} = \Phi_{\mathcal{x}}$ for convenience. From Eq.\eqref{moments1}, 

\begin{equation*}
\begin{split}
s_{n}^{(\mathcal{x})}
&= \Tr\left[
    \left(\sum_{i_1=1}^4 \Phi_{i_1}(\rho)\right)
    \left(\sum_{i_2=1}^4 \Phi_{i_2}(\rho)\right)
    \cdots
    \left(\sum_{i_n=1}^4 \Phi_{i_n}(\rho)\right)
\right] \\[6pt]
&= \Tr\!\left[
    \prod_{k=1}^n 
    \left( \sum_{i_k=1}^4 \Phi_{i_k}(\rho) \right)
\right] \\[6pt]
&= \Tr\!\left[
    \sum_{i_1, i_2, \ldots, i_n}
    \sum_{p_{k}^{(1)},\, p_{k}^{(2)},\, p_{k}^{(3)}}
    \prod_{k=1}^n \Phi_{i_k}(\rho)
\right. \\[4pt]
&\hspace{2cm}\left.
    \ket{p_{k}^{(1)} p_{k}^{(2)} p_{k}^{(3)}}
    \bra{p_{k}^{(1)} p_{k}^{(2)} p_{k}^{(3)}}
\right].
\end{split}
\end{equation*}

where $$I_{d^3} = \sum_{p_{k}^{(1)}=1}^2 \sum_{p_k^{(2)}=1}^2 \sum_{p_k^{(3)}=1}^2 \ket{p_{k}^{(1)} p_{k}^{(2)} p_{k}^{(3)}}\bra{p_{k}^{(1)} p_{k}^{(2)} p_{k}^{(3)}}$$ is the Identity operator. 

\begin{equation*}
\begin{split}
s_{n}^{(\mathcal{x})}
= & \sum_{i_1, i_2, \ldots, i_n}
    \Tr\!\Biggl[
        \prod_{k=1}^n \Phi_{i_k}(\rho)
        \sum_{p_{k}^{(1)},\, p_{k}^{(2)},\, p_{k}^{(3)}}
\\[4pt]
&\hspace{1.5cm}
        \ket{p_{k}^{(1)} p_{k}^{(2)} p_{k}^{(3)}}
        \bra{p_{k}^{(1)} p_{k}^{(2)} p_{k}^{(3)}}
    \Biggr].
\end{split}
\end{equation*}

For $i_k = 4, \Phi_4(\rho) = c_{\mathcal{x}} I_{8}$. The terms with $i_k=4$ reduce to some lower order moments. So, we consider $i_k \in \{1,2,3\}$ for each $k \in \{1,2,...,n\}$ henceforth. Considering one particular term in this expression gives
\begin{equation*}
\begin{split}
    s_{n}^{(\mathcal{x})} = & \Tr[\prod_{k=1}^n  \sum_{p_{k}^{(1)} p_{k}^{(2)} p_{k}^{(3)}} \Phi_{i_k}(\rho) \ket{p_{k}^{(1)} p_{k}^{(2)} p_{k}^{(3)}}\bra{p_{k}^{(1)} p_{k}^{(2)} p_{k}^{(3)}}] \\ & = \prod_{k=1}^n \sum_{p_{k}^{(1)} p_{k}^{(2)} p_{k}^{(3)}} \langle p_{k-1}^{(1)} p_{k-1}^{(2)} p_{k-1}^{(3)}|\Phi_{i_k}(\rho)| p_{k}^{(1)} p_{k}^{(2)} p_{k}^{(3)}\rangle
    \end{split}
\end{equation*}  where $p_0^{i_k} = p_n^{i_k}$ for $i_k \in \{1,2,3\}$. Note that considering one particular term fixes the value of the string $(i_1, i_2,...,i_n)$.

\begin{equation*}
\begin{split}
s_{n}^{(\mathcal{x})} 
= &\, \Tr\!\Biggl[
    \sum_{p_{1}^{(1)}, p_{1}^{(2)}, p_{1}^{(3)}, \ldots, p_{n}^{(1)}, p_{n}^{(2)}, p_{n}^{(3)}} 
    \prod_{k=1}^n \Phi_{i_k}(\rho)
\\[4pt]
&\hspace{2.2cm}
    \ket{p_{k}^{(1)} p_{k}^{(2)} p_{k}^{(3)}}
    \bra{p_{k-1}^{(1)} p_{k-1}^{(2)} p_{k-1}^{(3)}}
\Biggr]
\\[6pt]
= &\, \Tr\!\Biggl[
    \sum_{p_{1}^{(1)}, p_{1}^{(2)}, p_{1}^{(3)}, \ldots, p_{n}^{(1)}, p_{n}^{(2)}, p_{n}^{(3)}} 
    \prod_{k=1}^n \rho\, \Phi_{i_k}^{*}
\\[4pt]
&\hspace{2.2cm}
    \!\Bigl(\ket{p_{k}^{(1)} p_{k}^{(2)} p_{k}^{(3)}}
    \bra{p_{k-1}^{(1)} p_{k-1}^{(2)} p_{k-1}^{(3)}}
\Bigr)
\Biggr].
\end{split}
\end{equation*}

where $\Phi_{i_k}^{*}$ is dual to $\Phi_{i_k}$. The above expression reduces to

\begin{equation*}
\begin{split}
&\sum_{p_{1}^{(1)}, p_{1}^{(2)}, p_{1}^{(3)}, \ldots, p_{n}^{(1)}, p_{n}^{(2)}, p_{n}^{(3)}} 
\Tr\Biggl[
\rho^{\otimes n} 
\Bigl(
\otimes_{k=1}^n \Phi_{i_k}^{*} 
\\[4pt]
&\hspace{3.5cm}
\ket{p_{k}^{(1)} p_{k}^{(2)} p_{k}^{(3)}}
\bra{p_{k-1}^{(1)} p_{k-1}^{(2)} p_{k-1}^{(3)}}\Bigr)
\Biggr],
\end{split}
\end{equation*}

where we have used the properties that $\Tr[(A \otimes B)(C \otimes D)] = \Tr[AC]\Tr[BD]$, and $\Tr[A \otimes B] = \Tr[A]\Tr[B]$ for finite-dimensional matrices $A,B,C,D$, and $A(B)$ is of the same size as $C(D)$.
Defining $\Phi_{i_k}^{*} = \otimes_{j_{k}=1}^{3} \chi_{j_k}^{(i_k)}$ where
\begin{align*}
\chi_{j_k}^{(i_k)}=
\left\{
    \begin {aligned}
         & \Lambda_{\mathcal{x}}^* \quad & \text{for} \hspace{1mm} j_k = i_k \\
         & \mathbb{I} \quad & \text{otherwise,}                   
    \end{aligned}
\right.
\end{align*} and $\Lambda_{\mathcal{x}}^*$ is dual to $\Lambda_{\mathcal{x}}$. So,
\begin{equation*}
\begin{split}
    s_{n}^{(\mathcal{x})} = & \sum_{p_{1}^{(1)} p_{1}^{(2)} p_{1}^{(3)}...p_{n}^{(1)} p_{n}^{(2)} p_{n}^{(3)}} \Tr[ \rho^{\otimes n} (\otimes_{k=1}^{n} (\otimes_{j_{k}=1}^3 \chi_{j_k}^{(i_k)}\ket{p_k^{j_k}}\bra{p_k^{j_k}}))] \\ & = \Tr[\rho^{\otimes n}(O_{A_1 A_2...A_n} \otimes O_{B_1 B_2...B_n} \otimes O_{C_1 C_2...C_n})] 
    \end{split}
\end{equation*} where 
\begin{equation}O_{A_1 A_2...A_n} = (\sum_{p_1^{(1)}p_2^{(1)}...p_n^{(1)}}\otimes_{k=1}^{n}\chi_{1}^{(i_k)}\ket{p_k^{(1)}}\bra{p_{k-1}^{(1)}}), \label{op1}
\end{equation} 
\begin{equation}
O_{B_1 B_2...B_n} = (\sum_{p_1^{(2)}p_2^{(2)}...p_n^{(2)}}\otimes_{k=1}^{n}\chi_{2}^{(i_k)}\ket{p_k^{(2)}}\bra{p_{k-1}^{(2)}}),\label{op2}
\end{equation}
\begin{equation}
O_{C_1 C_2...C_n} = (\sum_{p_1^{(3)}p_2^{(3)}...p_n^{(3)}}\otimes_{k=1}^{n}\chi_{3}^{(i_k)}\ket{p_k^{(3)}}\bra{p_{k-1}^{(3)}}).\label{op3}
\end{equation}
For each term appearing in the expression of the moment, this procedure is followed to find the operators corresponding to that term. The $n$th order moment is obtained by finding the expectation values of these operators on $n$ copies of the state.

Eqs.\eqref{op1},\eqref{op2},\eqref{op3} represent the operators for any order of moment.

For a well-known map such as the transposition map ($\mathcal{x}=\mathcal{T}$), the operators are given by Eqs.\eqref{op1},\eqref{op2},\eqref{op3}, with
\begin{align*}
\chi_{j_k}^{(i_k)}=
\left\{
    \begin {aligned}
         & \Lambda_{\mathcal{T}} \quad & \text{for} \hspace{1mm} j_k = i_k \\
         & \mathbb{I} \quad & \text{otherwise}                   
    \end{aligned}
\right.
\end{align*} since the transposition map is self-dual.
As discussed above, the $n$th order moments are obtained by finding the expectation values of these operators on $n$ copies of the state. 
\end{itemize}
Extending beyond the tripartite case, the GME map for an $N$ qubit state is given by Eq.\eqref{map}, and the $n$th order moments corresponding to this map are given by Eq.\eqref{moments}
where $n \in \mathbb{N}$. As $N$ increases, the number of terms increases, and evaluating the exact operators for an arbitrary $N$ is difficult to obtain analytically. However, one can resort to computational methods to obtain the terms. Though we have considered $N=3$ and $d=2$ for convenience, our method discussed above can be extended in a similar fashion to evaluate the moments corresponding to an arbitrary $N$ and $d$. The main steps are outlined below.\\
(i) For an arbitrary $N$ qudit state, find the operators corresponding to the $n$th order moment using the formalism presented above.\\
(ii) Measurement of the corresponding operators on $n$ copies of the state give the moments.

It may be worth reemphasizing here that the method of detecting bipartite entanglement using moments has been explored earlier in the literature, ranging from entanglement detection in qubits \cite{ali2025detection}, to higher-dimensional systems \cite{wang2022operational,aggarwal2024entanglement,ali2025detection,nzrc-8yrt}. The use of moments in some other resource theories has also been studied \cite{PhysRevA.109.022247,PhysRevA.111.032406,wang2025notes}. However, since these methods pertain to bipartite systems, they are not directly comparable to our approach, which is specifically designed for multipartite systems.
\item In multipartite systems, Ref. \cite{ali2025detection} discusses the detection of $GHZ$ and $W$ states. However, their proposed criterion detects the NPT-ness of these states and is not formulated to detect the genuine entanglement in them.

On the other hand, it may be worthwhile to compare our approach with other methods suggested in the literature to detect GME in experiments. For instance, a criterion based on the norms of correlation vectors requires measuring all the generators of $SU(d)$~\cite{PhysRevA.96.052314}. Finding the expectation values of all such generators on the state corresponds to the full state tomography, leading
to an exponential growth in the number of measurements for multipartite systems.   In contrast, our moment-based criteria can be evaluated by using some suitable operators. Schemes based on
entanglement witnesses \cite{Zhou_2019, PhysRevA.107.052405}, rely on
prior knowledge of the state, since a single witness cannot detect all
states. For example, the construction of witnesses in \cite{Zhou_2019} is valid for graph states only. 
Device-independent methods to detect GME have also been suggested
\cite{PhysRevX.7.021042, PhysRevLett.122.060502}. 
Such methods turn out to be ineffective for general mixed states, 
except for states arbitrarily close to the pure genuine multipartite entangled states. In contrast, our approach has the potential for applicability to arbitrary mixed states,
as well. Further, device-independent approaches based on Bell-violation
are unable to detect genuine multipartite entangled states admitting local-hidden-variable models.
For example, the noisy $GHZ$ state, which admits a local-hidden-variable model for $\mu < 0.4688$ \cite{designolle2023improved}, is genuinely entangled for $\mu>0.428$, as observed through an optimized transposition map \cite{clivaz2017genuine}. It may be interesting to investigate the
efficacy of our truncated moment-based approach for possible detection
of other such examples of genuine multipartite entangled states admitting local-hidden-variable description.

\section{Conclusions}\label{s5}
Genuine multipartite entanglement (GME), the strongest form of entanglement in multipartite systems, is a resource \cite{cleve1997substituting,buhrman1999multiparty,puliyil2022thermodynamic,
sun2024genuine} offering significant advantages over its bipartite counterpart \cite{yamasaki2018multipartite,perseguers2013distribution,navascues2020genuine}. However, harnessing such advantages in practical tasks requires efficient detection of GME. 
Driven by the motivation to detect GME in an experimentally feasible way, we propose a method to detect GME based on truncated moments of positive maps. To validate our method, we demonstrate its effectiveness in detecting two inequivalent classes of tripartite entangled states, as well as four-qubit $GHZ$ state using our moment-based criterion implemented via the transposition, and the modified transposition map respectively. Additionally, we illustrate through explicit examples how moments of the reduction map can also serve as effective tools for detecting GME. 
 
Further, we propose an experimental scheme for evaluating the moments of the transposition map. A key advantage of our approach over traditional positive map-based criteria lies in its experimental feasibility: since such maps are unphysical, they cannot be directly implemented in practice.  On the contrary, our proposed moments are linear functionals that are realizable as expectation values of suitably chosen operators \cite{cieslinski2024analysing, PhysRevA.91.032315, PhysRevLett.121.150503}, and can be experimentally estimated using shadow tomography \cite{aaronson2018shadow}. Unlike full quantum state tomography, which demands an exponential number of state copies, the present approach enables our scheme to operate with only a polynomial number of state copies \cite{huang2020predicting}, and is applicable for both pure and mixed states.

 It may be noted that the required moment order to detect GME of an arbitrary state depends on the specific class and type of states under consideration, and there should not be any general bound. Considering that evaluating all the moments is equivalent to the full state tomography, as in the bipartite case \cite{neven2021symmetry}, our examples strongly motivate the use of moments as a practical tool for genuine entanglement detection: although the required moment order may in principle be arbitrarily high even for systems of low dimensions, it is finite in most of the known classes of states. Evaluating a finite number of moments is therefore efficient compared to the full state tomography. However, establishing exact bounds on the moment order for particular classes of states is left open for future investigation.  Note further, that the maps introduced here are decomposable, and hence are not suited for the detection of PPT entangled states. But, our method of moments is general in the sense that one can define moments corresponding to an indecomposable map and study its effectiveness in detecting  GME PPT entangled states.

Our study opens up several  other prospects for future research. While we have focused primarily on the transposition map, one can similarly define moments corresponding to other positive maps that may be able to detect a broader class of states or exhibit greater robustness to noise. Also, finding the explicit operator(s) corresponding to other positive maps and their experimental implementation is a natural offshoot of our present analysis. Further, as the order of moments increases, the detection criteria become more stringent, potentially leading to tighter bounds. This feature could be particularly advantageous for detecting wider categories of mixed states. It would also be interesting to apply our approach  to states admitting local-hidden-variable description.

\section{Acknowledgements}\label{s6}
B.M. acknowledges the DST INSPIRE fellowship program for financial support. S.G.N. acknowledges support from the CSIR project $09/0575(15951)/2022$-EMR-I.
\bibliography{main}
\onecolumngrid
\appendix

\section{Genuine entanglement detection using moments of reduction map} \label{A}
\subsection{\textbf{Detection of $GHZ$ state using the moments of reduction map:}}
Consider the tripartite $GHZ$ state, $\ket{GHZ ^{(3)}} \in \mathcal{D}(\mathbb{C}^{2} \otimes \mathbb{C}^{2} \otimes \mathbb{C}^{2})$ represented by
\begin{equation}
\ket{GHZ ^{(3)}}= \frac{\ket{000}+\ket{111}}{\sqrt{2}} . \label{GHZ}
\end{equation}  \\
We use the positive map given by Eq.~\eqref{map} for the detection of genuine entanglement of the above state with $\mathcal{x} = \mathcal{R}$, where $\Lambda_{\mathcal{R}}$ is the qubit reduction map defined in Eq.~\eqref{reductionmap} and $c_{\mathcal{R}}^{(3)} = 1$

In fact, applying the map $\Phi^{(3)}_{\mathcal{R}}$ to the state $\ket{\mathrm{GHZ}^{(3)}}\bra{\mathrm{GHZ}^{(3)}}$ yields a negative value, indicating that the state is genuinely entangled. The corresponding moments for this map can be computed using Eq.~\eqref{moments}. Utilizing moments up to the third and fifth order, we obtain $\det[H_1(\mathbf{s^{(\mathcal{R})}})] < 0$ and $\det[H_2(\mathbf{s^{(\mathcal{R})}})] < 0$. These results demonstrate that the moments associated with the reduction map can be effectively used to detect the genuine entanglement of the $GHZ$ state.\\

\subsection{\textbf{Detection of noisy $GHZ$ state using moments of reduction map:}}
\begin{figure}[ht]
\includegraphics[width=0.5\textwidth]{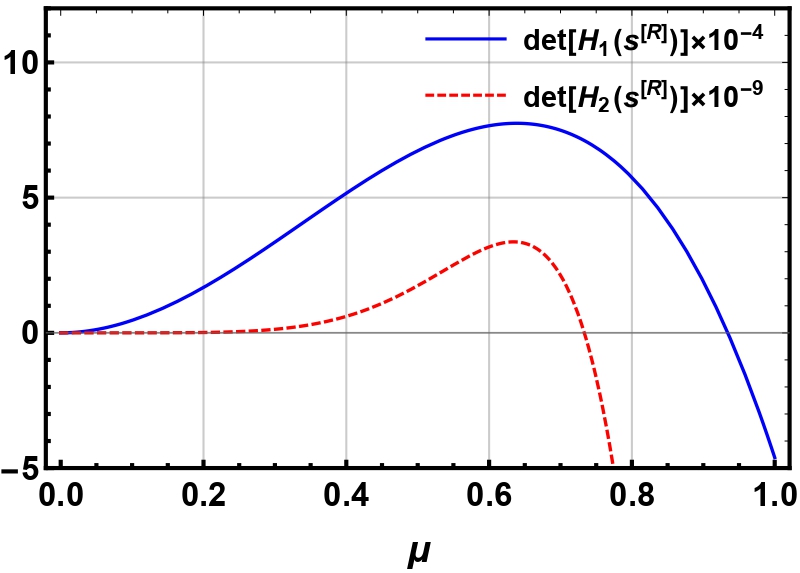}
\caption{Detecting genuine entanglement of tripartite noisy $GHZ$ state using the moments of reduction map}\label{fig4}
\centering
\end{figure}

Consider the noisy $GHZ$ state represented by 
\begin{equation}
    \ket{GHZ ^{(3)}}_\mu\bra{GHZ ^{(3)}} = \mu \ket{GHZ ^{(3)}}\bra{GHZ ^{(3)}} + \frac{(1-\mu)}{8} I_{8} \label{noisyGHZ}
\end{equation}
where $\mu \in [0, 1]$ is the noise parameter.
Here $\Phi^{(3)}_{\mathcal{R}}(\ket{GHZ ^{(3)}}_\mu\bra{GHZ ^{(3)}})$ has a negative eigen value for $\mu > 0.733$ whereas $\det[H_1(\mathbf{s}^{(\mathcal{R})})]$ is negative for $\mu > 0.934$. This suggests that the first Hankel matrix condition is not sufficient to detect the genuine entanglement of this state for $0.733 < \mu \le 0.934$. However, one may use the conditions involving higher order moments. For e.g., $\det[H_2(\mathbf{s}^{(\mathcal{R})})] < 0$ for $\mu > 0.733$, indicating that violation of the second Hankel matrix criterion (which involves moments up to \textit{fifth} order) is sufficient to detect the genuine entanglement for the entire range. Fig.~\ref{fig4} shows the violation of the first and second Hankel matrix conditions with the noise parameter.\\

\subsection{\textbf{Detection of convex mixture of $GHZ$ and $W$ state using moments of reduction map:}}
\begin{figure}[h]
\includegraphics[width=0.52\textwidth]{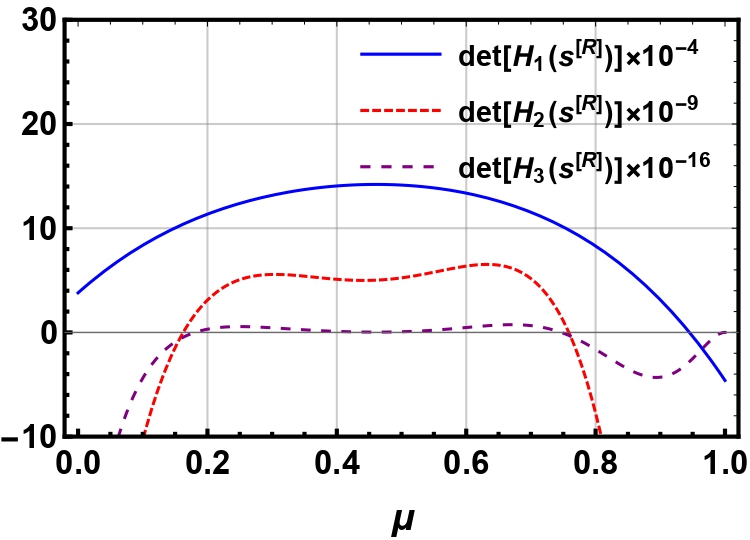}
 \caption{Detection of genuine entanglement of $\ket{\psi^{(3)}}\bra{\psi^{(3)}}$ using the moments of reduction map}\label{fig5}
\centering
\end{figure}

Let $\ket{\psi^{(3)}}\bra{\psi^{(3)}}$ be the tripartite state $\in \mathcal{D}(\mathbb{C}^{2} \otimes \mathbb{C}^{2} \otimes \mathbb{C}^{2})$, represented by
\begin{equation}
    \ket{\psi^{(3)}}\bra{\psi^{(3)}} = \mu \ket{GHZ^{(3)}}\bra{GHZ^{(3)}} + (1-\mu) \ket{W^{(3)}}\bra{W^{(3)}}.
\end{equation}  \label{convexmixture}
Using the reduction map, the state $\ket{\psi^{(3)}}\bra{\psi^{(3)}}$ exhibits negative eigenvalues for $\mu < 0.182$ and $\mu > 0.746$, indicating entanglement in these regions. However, the first Hankel matrix criterion detects genuine entanglement only for $\mu > 0.945$. When higher order moments are considered, the second Hankel matrix condition identifies genuine entanglement for $\mu \leq 0.162$ and $\mu > 0.758$, whereas the entire range of genuine entanglement is detected using the third Hankel matrix criterion. So, the map-based criterion and our moment-based criterion are equivalent for moments up to \textit{seventh} order. The corresponding violation of the first, second, and third Hankel matrix criteria as a function of the noise parameter $\mu$ is depicted in Fig.~\ref{fig5}.

\section{Evaluation of the moments used in our analysis} \label{B}
We provide the detailed calculation for the terms appearing in the expressions of the second and third order moments. This can be done for all the other higher order moment terms in a similar fashion.
\subsection{Expressions for the terms in the second order moment } \label{B1}
In the expression for the second order moment, there are two types of terms (Eq.~\eqref{secondordermomentsfinal}). All the terms can be evaluated using
the following calculations. 

$\bullet$ \textit{Calculation of $\Tr[(\Phi_1 (\rho))^2] : $}
 $$\Tr[(\Phi_1 (\rho))^2] =  \sum_{pqrijk} \bra{pqr} \rho_{ABC}^{\Lambda_{\mathcal{T}_{A}}} \ket{ijk} \bra{ijk} \rho_{ABC}^{\Lambda_{\mathcal{T}_{A}}} \ket{pqr}$$  $$= \sum_{pqrijk} \bra{iqr} \rho_{ABC} \ket{pjk} \bra{pjk} \rho_{ABC} \ket{iqr} $$ $$= \sum_{pqrijk} \Tr[(\rho_{ABC} \otimes \rho_{A^{'} B^{'} C^{'}})  (\ket{pjk}_{A B C} \bra{iqr} \otimes \ket{iqr}_{A^{'} B^{'} C^{'}} \bra{pjk})]$$ $$ = \Tr[(\rho_{ABC} \otimes \rho_{A^{'} B^{'} C^{'}})(SWAP_{A A^{'}} \otimes SWAP_{B B^{'}} \otimes SWAP_{C C^{'}})]$$
 Therefore, for finding the second order moment, two copies of the unknown state are given to the parties Alice, Bob, and Charlie. The parties find the expectation value of $\otimes_{u \in \{A,B,C\}} SWAP_{u u^{'}}$ to realize this term in experiments.
 
$\bullet$  \textit{Calculation of $\Tr[(\Phi_1 (\rho) \Phi_2 (\rho) )] : $}
 $$\Tr[(\Phi_1 (\rho) \Phi_2 (\rho))] =  \sum_{pqrijk} \bra{pqr} \rho_{ABC}^{\Lambda_{\mathcal{T}_{A}}} \ket{ijk} \bra{ijk} \rho_{ABC}^{\Lambda_{\mathcal{T}_{B}}} \ket{pqr}$$  $$= \sum_{pqrijk} \bra{iqr} \rho_{ABC} \ket{pjk} \bra{iqk} \rho_{ABC} \ket{pjr} $$ $$= \sum_{pqrijk} \Tr[(\rho_{ABC} \otimes \rho_{A^{'} B^{'} C^{'}})  (\ket{pjk}_{A B C} \bra{iqr} \otimes \ket{pjr}_{A^{'} B^{'} C^{'}} \bra{iqk})]$$ $$ = \Tr[(\rho_{ABC} \otimes \rho_{A^{'} B^{'} C^{'}})(2 \ket{\phi^+} _{A A^{'}}\bra{\phi^+} \otimes 2 \ket{\phi^+} _{B B^{'}}\bra{\phi^+} \otimes SWAP_{C C^{'}})]$$
 This is the expectation value of $\hat{\phi}_{A A^{'}} \otimes \hat{\phi}_{B B^{'}} \otimes SWAP_{C C^{'}}$ (with the corresponding operators defined in Eq.~\eqref{SWAP} and Eq.~\eqref{Y}) on two copies of the state.

    \subsection{Expressions for the terms in the  third order moment} \label{B2}
    Here, we present the detailed calculation for the typical terms appearing in Eq.~\eqref{thirdordermomentsfinal}. Using these expressions,
    all the other terms can be easily obtained. Each of these terms is the expectation value of some operators, as listed below. 
     
    \begin{equation}
        S^{b}_u = (\text{SWAP}_{u_1 u_2} \otimes I_{u_3}) (I_{u_1} \otimes \text{SWAP}_{u_2 u_3})  \label{backwardswap}
    \end{equation}
    
    \begin{equation}
        S^{f}_u = (SWAP_{u_1 u_3} \otimes I_{u_2}) (I_{u_1} \otimes SWAP_{u_2 u_3})  \label{forwardswap}
    \end{equation}

    \begin{equation}
      X_u = (\hat{\phi}_{u_1 u_{2}} \otimes I_{u_3}) (I_{u_1} \otimes \hat{\phi}_{u_2 u_3})  \label{operatorx}
    \end{equation}

    \begin{equation}
      Y_u = (\hat{\phi}_{u_1 u_3} \otimes I_{u_2}) (I_{u_1} \otimes SWAP_{u_2 u_3})  \label{operatory}
    \end{equation}

     \begin{equation}
      Z_u = (I_{B_1}  \otimes \hat{\phi}_{B_2 B_{3}} ) (\hat{\phi}_{B_1 B_3} \otimes I_{B_2})   \label{operatorz}
    \end{equation}

for $u \in \{A,B,C\}$. $S^{b}$ and $S^{f}$ represent the backward and forward $SWAP$ operators respectively.

   $\bullet$  \textit{Calculation of $\Tr[(\Phi_1 (\rho))^3] : $}
    $$\Tr[(\Phi_1 (\rho))^3] = \sum_{pqrijklmn} \bra{pqr} \rho_{ABC}^{\Lambda_{\mathcal{T}_A}} \ket{ijk} \bra{ijk} \rho_{ABC}^{\Lambda_{\mathcal{T}_A}} \ket{lmn} \bra{lmn} \rho_{ABC}^{\Lambda_{\mathcal{T}_A}} \ket{pqr}$$  $$ = \sum_{pqrijklmn}  \bra{iqr} \rho_{ABC} \ket{pjk} \bra{ljk} \rho_{ABC} \ket{imn} \bra{pmn} \rho_{ABC} \ket{lqr}$$ $$=\sum_{pqrijklmn} \Tr[(\rho_{A_1 B_1 C_1} \otimes \rho_{A_2 B_2 C_2} \otimes \rho_{A_3 B_3 C_3})  (\ket{pjk}_{A_1 B_1 C_1} \bra{iqr} \otimes \ket{imn}_{A_2 B_2 C_2} \bra{ljk} \otimes \ket{lqr}_{A_3 B_3 C_3} \bra{pmn})]$$ $$ = \Tr[(\rho_{A_1 B_1 C_1} \otimes \rho_{A_2 B_2 C_2} \otimes \rho_{A_3 B_3 C_3})(S^{b}_{A} \otimes S^{f}_{B} \otimes S^{f}_{C})]$$
where $S^{b}$ and $S^{f}$ are represented by Eqs.~\eqref{backwardswap} and \eqref{forwardswap} respectively. 

 $\bullet$ \textit{Calculation of $\Tr[(\Phi_1 (\rho))^2 \Phi_2(\rho)] : $}
 $$\Tr[(\Phi_1 (\rho))^2 \Phi_2(\rho)] = \sum_{pqrijklmn} \bra{pqr} \rho_{ABC}^{\Lambda_{\mathcal{T}_A}} \ket{ijk} \bra{ijk} \rho_{ABC}^{\Lambda_{\mathcal{T}_A}} \ket{lmn} \bra{lmn} \rho_{ABC}^{\Lambda_{\mathcal{T}_B}} \ket{pqr}$$ $$ = \sum_{pqrijklmn}  \bra{iqr} \rho_{ABC} \ket{pjk} \bra{ljk} \rho_{ABC} \ket{imn} \bra{lqn} \rho_{ABC} \ket{pmr}$$ $$=\sum_{pqrijklmn} \Tr[(\rho_{A_1 B_1 C_1} \otimes \rho_{A_2 B_2 C_2} \otimes \rho_{A_3 B_3 C_3})  (\ket{pjk}_{A_1 B_1 C_1} \bra{iqr} \otimes \ket{imn}_{A_2 B_2 C_2} \bra{ljk} \otimes \ket{pmr}_{A_3 B_3 C_3} \bra{lqn})]$$ $$ = \Tr[(\rho_{A_1 B_1 C_1} \otimes \rho_{A_2 B_2 C_2} \otimes \rho_{A_3 B_3 C_3})(S^{b}_{A} \otimes Z_B \otimes S^{f}_{C})]$$
 The corresponding operators are defined in Eqs.~\eqref{backwardswap}, \eqref{operatorz}, and \eqref{forwardswap} respectively.

  $\bullet$  \textit{Calculation of $\Tr[\Phi_1 (\rho) \Phi_2(\rho) \Phi_3(\rho)] : $} 
  
  This can be realized as follows
  $$\Tr[\Phi_1 (\rho) \Phi_2(\rho)  \Phi_3(\rho)] = \sum_{pqrijklmn} \bra{pqr} \rho_{ABC}^{\Lambda_{\mathcal{T}_A}} \ket{ijk} \bra{ijk} \rho_{ABC}^{\Lambda_{\mathcal{T}_B}} \ket{lmn} \bra{lmn} \rho_{ABC}^{\Lambda_{\mathcal{T}_C}} \ket{pqr}$$ $$ = \sum_{pqrijklmn}  \bra{iqr} \rho_{ABC} \ket{pjk} \bra{imk} \rho_{ABC} \ket{ljn} \bra{lmr} \rho_{ABC} \ket{pqn}$$ $$=\sum_{pqrijklmn} \Tr[(\rho_{A_1 B_1 C_1} \otimes \rho_{A_2 B_2 C_2} \otimes \rho_{A_3 B_3 C_3})  (\ket{pjk}_{A_1 B_1 C_1} \bra{iqr} \otimes \ket{ljn}_{A_2 B_2 C_2} \bra{imk} \otimes \ket{pqn}_{A_3 B_3 C_3} \bra{lmr})]$$ $$ = \Tr[(\rho_{A_1 B_1 C_1} \otimes \rho_{A_2 B_2 C_2} \otimes \rho_{A_3 B_3 C_3})(Y_A \otimes X_B \otimes Z_C)].$$
 where $X_B, Y_A$ and $Z_C$ are represented in Eqs.~\eqref{operatorx}, \eqref{operatory} and \eqref{operatorz} respectively.
  
  The expectation values of the respective operators on three copies of the state would enable experimental evaluation of these terms. 

\end{document}